\documentclass[lettersize,journal]{IEEEtran}
\usepackage{amsmath,amssymb,mathrsfs,amsfonts}
\usepackage{algorithmic}
\usepackage{algorithm}
\usepackage{array}
\usepackage[caption=false,font=footnotesize,labelfont=rm,textfont=rm]{subfig}
\usepackage{textcomp}
\usepackage{stfloats}
\usepackage{url}
\usepackage{verbatim}
\usepackage{graphicx}
\usepackage{cite}
\usepackage[square, comma, sort&compress, numbers]{natbib}
\usepackage{multirow}
\usepackage{makecell}
\usepackage{diagbox}
\usepackage{subfig}
\usepackage{subfloat}
\usepackage{color}
\newcommand{\mathbbm}[1]{\text{\usefont{U}{bbm}{m}{n}#1}}

\begin{document}

\title{AI-Native 6G Physical Layer with Cross-Module Optimization and Cooperative Control Agents}

\author{Xufei Zheng, Han Xiao, Shi Jin, Zhiqin Wang, Wenqiang Tian, Wendong Liu, Jianfei Cao, Jia Shen, Zhihua Shi, Zhi Zhang, and Ning Yang

\thanks{X. Zheng, H. Xiao, W. Tian, W. Liu, J. Cao, J. Shen, Z. Shi, Z. Zhang, and N. Yang are with the Dept. of Standardization, OPPO Research Institute, Beijing, China (e-mail: zhengxufei@oppo.com; xiaohan1@oppo.com; tianwenqiang@oppo.com; liuwendong1@oppo.com; caojianfei@oppo.com; sj@oppo.com; szh@oppo.com; zhangzhi@oppo.com; yangning@oppo.com).} 

\thanks{S. Jin is with the National Mobile Communications
Research Laboratory, Southeast University, Nanjing, China
(e-mail: jinshi@seu.edu.cn) ({\it Corresponding author: Shi Jin}).} 

\thanks{Z. Wang is with the China Academy of Information and Communications Technology, Beijing, China
(e-mail: zhiqin.wang@caict.ac.cn).}


}



\maketitle

\begin{abstract}
In this article, a framework of artificial intelligence (AI)-native cross-module optimized physical layer with cooperative control agents is proposed, which involves optimization across global AI/machine learning (ML) modules of the physical layer with innovative design of multiple enhancement mechanisms and control strategies. Specifically, it achieves simultaneous optimization across global modules of uplink AI/ML-based joint source-channel coding with modulation, and downlink AI/ML-based modulation with precoding and corresponding data detection, reducing traditional inter-module information barriers to facilitate end-to-end optimization toward global objectives. Moreover, multiple enhancement mechanisms are also proposed, including i) an AI/ML-based cross-layer modulation approach with theoretical analysis for downlink transmission that breaks the isolation of inter-layer features to expand the solution space for determining improved constellation, ii) a utility-oriented precoder construction method that shifts the role of the AI/ML-based CSI feedback decoder from recovering the original CSI to directly generating precoding matrices aiming to improve end-to-end performance, and iii) incorporating modulation into AI/ML-based CSI feedback to bypass bit-level bottlenecks that introduce quantization errors, non-differentiable gradients, and limitations in constellation solution spaces. Furthermore, AI/ML-based control agents for optimized transmission schemes are proposed that leverage AI/ML to perform model switching according to channel state, thereby enabling integrated control for global throughput optimization. Finally, simulation results demonstrate the superiority of the proposed solutions in terms of block error rate and throughput. These extensive simulations employ more practical assumptions that are aligned with the requirements of the 3rd Generation Partnership Project (3GPP), which hopefully provides valuable insights for future 3GPP standardization discussions.
\end{abstract}


\section{Introduction}
\IEEEPARstart{T}{he} generational evolution of wireless communication systems is inherently intertwined with the deep integration of transformative technologies, especially a series of innovations in the physical layer \cite{series2015imt,5Ghai,TS38913}. Specifically, due to the successful application of artificial intelligence (AI) in the field of computer vision and natural language processing, the enhancement of wireless communication using AI has attracted great attention in recent years \cite{huang2019deep}, where the data-driven nature and nonlinear processing capabilities of AI solutions bring more adaptability and performance gains compared to other traditional solutions. 
A series of creative academic works have been done one after another, such as channel state information (CSI) feedback \cite{han2024ai, xiao2021ai, liu2021evcsinet, bi2022novel}, channel estimation \cite{le2021deep, gao2023deep}, beamforming \cite{al2022review}, modulation \cite{alberge2018deep, jiang2019deep} and so on. The robustness of channel estimation under imperfect CSI conditions has been further extensively studied in the fifth generation (5G) systems \cite{pourkabirian2021robust}, while emerging sixth generation (6G) technologies are pushing the boundaries of ultra-reliable low-latency communications through novel physical-layer innovations \cite{pourkabirian2024vision}. For the standardization process of 5G-Advanced, the 3rd Generation Partnership Project (3GPP) has also initiated studies on AI/machine learning (ML) for new radio (NR) air interface enhancements \cite{213599}, focusing on three pivotal use cases: CSI feedback enhancement, beam management, and high-precision positioning. These studies encompass evaluation methodologies, potential specification impacts, and other relevant aspects. Specifically, regarding CSI feedback enhancement, 3GPP Release 18 and Release 19 standardization efforts have centered on two technical directions: AI/ML-based CSI compression and AI/ML-based user equipment (UE)-side CSI prediction. These efforts aim to surpass the performance limitations of conventional codebook-based CSI feedback mechanisms \cite{TS0000,TS0001,TS0002,TS0003} while simultaneously reducing overhead and latency. For AI/ML-based beam management, the standardization focuses on spatial and temporal domain beam prediction to achieve reduced reference signal overhead and real-time beam information acquisition in high frequency bands. As for high-precision positioning, 3GPP aims to leverage AI/ML to significantly improve localization accuracy in non-line-of-sight (NLoS)-dominated environments. 

However, the above works in academia and industry still adopt a modular architecture, which decouples the physical layer into separate functional blocks and optimizes them independently. Although this compartmentalized design philosophy is beneficial for finding the optimal solution of independent modules and achieving engineering tractability, its inherent limitation, i.e., the globally suboptimal performance caused by module isolation, has become increasingly evident. 

Fortunately, with the introduction of AI/ML solutions, holistic design beyond existing modular architectures has become possible. The 6G presents an opportunity to revisit the fundamentals of radio system design and explore native integration with AI/ML, targeting transformative leaps in capability, efficiency and simplicity, which requires a holistic and optimized design across all layers of the radio interface and architecture \cite{251881, jiang2021road, yang20226g, hoydis2021toward, jiao2024advanced, ozpoyraz2022deep, abd2024evolution, salh2021survey}. Related works demonstrate a paradigm shift in research focus from module-level AI/ML features to systematic joint design. At the transmitter, joint design of the modulation and precoding \cite{zhang2023multi}, joint source-channel coding (JSCC) \cite{xu2022deep, chu2023deep,jiang2024deep} and modulation \cite{matsumoto2023impact,inoue2021deep} are studied. At the receiver, \cite{ait2021end} considers the joint design of channel estimation and data detection, realizing an integrated receiver by end-to-end training procedure. As for the downlink transmission with uplink feedback procedure, existing works propose joint CSI feedback and precoding \cite{sun2022deep,sohrabi2021deep, guo2024deep} with pilot design \cite{jang2022deep,jang2019deep} or channel estimation \cite{guo2025deep, zhao2022joint}. 

However, although the existing works achieve joint design of partial modules of the link, they still fail to consider more thorough exploration of the global degrees of freedom for design of physical layer with AI/ML features, i.e., i) simultaneous optimization across global modules of uplink AI/ML-based joint source-channel coding with modulation, and downlink AI/ML-based modulation with precoding and corresponding data detection, and ii) joint exploitation of the features across transmitting layers to expand the solution space for finding the optimal constellation and precoder. Moreover, existing works largely rely on idealized system assumptions, e.g., additive white Gaussian noise (AWGN) downlink channel or ideal uplink feedback, may encounter intractable challenges of adaptability when scaling to practical scenarios involving more practical channel model and imperfect uplink channel for CSI feedback.

In addition, both AI/ML-based and conventional wireless communication systems require demand-aware control strategies to dynamically adjust transmission schemes according to real-time channel conditions. However, there are still several reasons why we need to go beyond existing solutions and design new control strategies based on AI/ML. 
\begin{itemize}
\item 
Reviewing the existing communication systems, current implementations employ adaptive modulation and coding (AMC) control loops that perform reactive scheme updates based on expected performance. The granularity of available transmission schemes and their environmental sensitivity significantly impact AMC strategies, frequently necessitating trade-offs between implementation complexity and optimal scheme selection. However, with 3GPP Release 18 standardization initiating the integration of AI/ML-based solutions and the anticipated proliferation of wireless AI/ML technologies in 6G, conventional AMC mechanisms face emerging challenges. Specifically, traditional link adaptation mechanism utilizes a lookup table mapping channel quality index (CQI) to modulation and coding scheme (MCS), which is based on the block error rate (BLER) threshold derived from traditional non-AI/ML-based system. Once AI/ML features are introduced in the system and implementation form of link adaptation changes from MCS selection to AI/ML model switching, establishing mapping rules becomes challenging through theoretical derivation. This is attributed to the expanded number of candidate AI/ML schemes and the proprietary nature of these models, since neither UE nor network (NW) vendors typically disclose detailed transmission schemes.

\item 
Considering current state-of-the-art study, the model lifecycle management (LCM) \cite{chen20235g} in 3GPP studies alleviates the above difficulties through performance monitoring, model switching, and model updating procedure. However, it still poses a series of challenges. First, existing LCM solutions employ post-hoc adaptation with lower real-time performance, as switching is triggered only after performance degradation is detected during model monitoring. Besides, since more and more features with two-sided models are anticipated to be integrated into 6G systems, the existing LCM solutions will inevitably incur substantial signaling overhead and impose heavy standardization burdens, where each new two-sided feature requires additional control signaling over the air interface to ensure model alignment. 
\end{itemize}
These challenges necessitate novel design of control strategies for AI/ML-based system.

In this article, an AI-native integrated solution, namely AI-native cross-module optimized physical layer with cooperative control agents (CMO-CCA), is proposed, which involves optimization across global AI/ML modules of the link with innovative design of multiple enhancement mechanisms and control strategies. Specifically, the main contributions of this article are summarized as follows.

\begin{itemize}
\item 
An AI-native framework is proposed, which involves i) AI/ML-based control strategy capable of dynamically selecting the optimal transmission scheme based on real-time wireless channel conditions, while establishing two-sided model alignment between UE and NW, and ii) simultaneous optimization across global modules of uplink AI/ML-based joint source-channel coding with modulation, and downlink AI/ML-based modulation with precoding and corresponding data detection, breaking down traditional inter-module information silos to enable end-to-end optimization under global objectives.
\item 
Multiple enhancement mechanisms are proposed including i) an AI/ML-based cross-layer modulation approach with theoretical analysis for downlink transmission that breaks the isolation of inter-layer features to expand the solution space for determining optimal constellation, ii) a utility-oriented precoder construction method that shifts the role of the AI/ML-based CSI feedback decoder from recovering the original CSI to directly generating precoding matrices optimized for end-to-end performance, and iii) incorporating modulation into AI/ML-based CSI feedback to bypass bit-level bottlenecks that introduce quantization errors, non-differentiable gradients, and limitations in constellation solution spaces.
\item 
AI/ML-based control agents for optimized transmission schemes are proposed that leverage AI/ML to perform model switching according to channel state, thereby enabling integrated control for global throughput optimization.
\item 
Various kinds of simulation results are provided to demonstrate the superiority of the proposed solutions in terms of BLER and throughput. These extensive simulations employ more practical 3GPP wireless channel assumptions considering imperfect uplink feedback channel, which hopefully provides valuable insights for future 3GPP standardization discussions.
\end{itemize}

From the perspective of AI methodology, this work presents contributions across three key aspects. First, our novel end-to-end optimization framework introduces unique objective functions and constraints that address critical gaps in current AI approaches, substantially advancing the application of deep learning in communication systems. Second, we develop groundbreaking neural architectures featuring: i) a cross-layer modulation model enabling joint optimization of high-dimensional constellation mappings, ii) a novel quantization-free transformer-based CSI processing model eliminating quantization bottlenecks, and iii) a control model achieving alignment strategies. Third, we develop an advanced three-stage progressive learning strategy with full-link differentiability and dynamic loss weighting for stable optimization of coupled modules.

The remainder of this article is organized as follows. In section \ref{sectionII}, the system description is introduced including multi-input multi-output (MIMO) system with orthogonal frequency division multiplexing (OFDM) waveform, downlink data transmission and uplink CSI feedback. In section \ref{sectionIII}, the proposed CMO-CCA is introduced. Simulation results are provided in section \ref{SectionIV}. Standardization discussion is also provided in section \ref{sectionV_sub}. Final conclusions are given in section \ref{sectionV}.

Throughout this paper, upper-case and lower-case letters denote scalars. Boldface upper-case and boldface lower-case letters denote matrices and vectors, respectively. Calligraphic upper-case letters denote sets. Specifically, the sets of real and complex numbers are denoted by $\mathbb{R}$ and $\mathbb{C}$, respectively. To further enhance the readability and accessibility of the manuscript, a table summarizing the key mathematical notations with their definitions can be found in Table \ref{key_notations}.

\begin{table}[t]
\centering
\caption{Summary of Key Mathematical Notations}
\label{key_notations}
\begin{tabular}{|c|l|}
\hline
Notation & Definition \\ \hline
$\mathbf{b}$ & Original information bit stream \\ \hline
$\hat{\mathbf{b}}$ & Received information bit stream \\ \hline
$\mathbf{c}$ & Encoded bit stream \\ \hline
$\mathbf{s}$ & Modulated complex symbols \\ \hline
$\widetilde{\mathbb{C}}$ & Constellation set \\ \hline
$m$ & Modulation order \\ \hline
$\mathbf{y}_{f,t}$ & Received signal \\ \hline
$\mathbf{H}_{f,t}$ & Downlink channel matrix\\ \hline
$\mathbf{x}_{f,t}$ & Precoded transmit signal\\ \hline
$\mathbf{n}_{f,t}$ & Additive white Gaussian noise \\ \hline
$N_{\mathrm{tx}}$ & Number of transmit antennas \\ \hline
$N_{\mathrm{rx}}$ & Number of receive antennas \\ \hline
$N_{\mathrm{sc}}$ & Number of subcarriers \\ \hline
$N_{\mathrm{t}}$ & Number of OFDM symbols \\ \hline
$N_{\mathrm{sb}}$ & Number of subbands \\ \hline
$N_{\mathrm{layer}}$ & Number of transmission layers \\ \hline
$\mathbf{P}$ & Precoding matrix \\ \hline
$\mathbf{H}_{\mathrm{eq}}$ & Equivalent channel matrix \\ \hline
$\mathbf{R}_k$ & Spatial covariance matrix for subband $k$ \\ \hline
$\mathbf{v}_k$ & Eigenvectors matrix for subband $k$ \\ \hline
$\mathbf{\Lambda}_k$ & Eigenvalues matrix for subband $k$ \\ \hline
$\mathbf{W}$ & CSI matrix for feedback \\ \hline
$f_{\mathrm{e}}(\cdot)$ & CSI encoder function \\ \hline
$f_{\mathrm{d}}(\cdot)$ & CSI decoder function \\ \hline
$\Theta_{\mathrm{E}}, \Theta_{\mathrm{D}}$ & Trainable parameters for encoder/decoder \\ \hline
$\rho(\cdot)$ & Squared generalized cosine similarity \\ \hline
$g_{\mathrm{mod}}(\cdot)$ & AI/ML-based cross-layer modulator \\ \hline
$g_{\mathrm{demod}}(\cdot)$ & AI/ML-based cross-layer demodulator \\ \hline
$\Theta_{\mathrm{mod}}, \Theta_{\mathrm{demod}}$ & Trainable parameters for modulator and demodulator \\ \hline
$\mathbf{s}_{\mathrm{CSI}}$ & CSI feedback symbols \\ \hline
$g_{\mathrm{la}}(\cdot)$ & Control agent switching model \\ \hline
$\Phi$ & Trainable parameters for control agent \\ \hline
$\mathfrak{F}$ & Physical layer link configuration \\ \hline
$\mathfrak{M}_q$ & Modulation scheme candidate $q$ \\ \hline
$\mathfrak{P}_v$ & Precoding scheme candidate $v$ \\ \hline
$\mathfrak{C}_z$ & CSI feedback scheme candidate $z$ \\ \hline
$\mathfrak{G}(\cdot)$ & Control agent function \\ \hline
\end{tabular}
\end{table}

\section{System Description}\label{sectionII}
\subsection{MIMO-OFDM Systems}
A typical single-user MIMO-OFDM system operating in frequency division duplexing (FDD) mode with $N_{\mathrm{tx}}$ transmit antennas at base station (BS) and $N_{\mathrm{rx}}$ receive antennas at UE is assumed. Note that while time division duplexing (TDD) systems with reciprocity-based beamforming offer advantages in massive MIMO scenarios by eliminating downlink training overhead, the FDD framework is adopted in this work due to its compatibility with dominant 5G NR deployments. The proposed AI-native enhancements to CSI feedback remain relevant even in hybrid TDD/FDD systems where feedback is still required. The study considers signal transmission based on OFDM frame structure with $N_{\mathrm{sc}}$ subcarriers and $N_{\mathrm{t}}$ OFDM symbols, where the subcarrier index and OFDM symbol index are represented by $f\in\{1,\ldots,N_{\mathrm{sc}}\}$ and $t\in\{1,\ldots,N_{\mathrm{t}}\}$, respectively.  Given a sufficiently long cyclic prefix to mitigate inter-symbol interference, the baseband-equivalent MIMO transmission process for each temporal-frequency resource element (RE) in the OFDM grid can be expressed as,
\begin{equation}\label{eq1}
\mathbf{y}_{f,t} = \mathbf{H}_{f,t}\mathbf{x}_{f,t}+\mathbf{n}_{f,t}
\end{equation}
where $\mathbf{y}_{f,t}\in\mathbb{C}^{N_{\mathrm{rx}}\times 1}$ is the received signal, $\mathbf{H}_{f,t}\in\mathbb{C}^{N_{\mathrm{rx}}\times N_{\mathrm{tx}}}$ is the downlink channel matrix, 
$\mathbf{x}_{f,t}\in\mathbb{C}^{N_{\mathrm{tx}}\times 1}$ is the precoded transmit signal and 
$\mathbf{n}_{f,t}\in\mathbb{C}^{N_{\mathrm{rx}}\times 1}$ is the additive white Gaussian noise with variance $\sigma_{n}^{2}$. Correspondingly, $\mathbf{y}\in\mathbb{C}^{N_{\mathrm{rx}}\times N_{\mathrm{sc}} \times N_{\mathrm{t}}}$, 
$\mathbf{H}\in\mathbb{C}^{N_{\mathrm{rx}}\times N_{\mathrm{tx}} \times N_{\mathrm{sc}} \times N_{\mathrm{t}}}$, 
$\mathbf{x}\in\mathbb{C}^{N_{\mathrm{tx}}\times N_{\mathrm{sc}} \times N_{\mathrm{t}}}$ and 
$\mathbf{n}\in\mathbb{C}^{N_{\mathrm{rx}}\times N_{\mathrm{sc}} \times N_{\mathrm{t}}}$ represent the received signal, downlink channel, precoded transmit signal and noise on all subcarriers and OFDM symbols, respectively. Based on the above system, data transmission and CSI feedback via downlink and uplink channel respectively can be achieved.

\subsection{Data Transmission via Downlink Channel}
At the BS side, the downlink processing starts with the encoding of information bit stream 
$\mathbf{b}\in\{0, 1\}^{N_{\mathrm{b}}\times 1}$ using a 5G-compliant low-density parity-check (LDPC) channel encoder, resulting in encoded bit stream
$\mathbf{c}\in\{0, 1\}^{N_{\mathrm{c}}\times 1}$. The coded bit stream undergoes modulation to form complex symbols
$\mathbf{s}\in\widetilde{\mathbb{C}}^{N_{\mathrm{s}}\times 1}$, where $\widetilde{\mathbb{C}}$ denotes the constellation set and the symbol sequence length satisfies $N_{\mathrm{s}}=N_{\mathrm{c}}/m$ with $m$ denoting the modulation order. The modulated symbols with inserted demodulation reference signals (DMRS), undergo precoding through multiplication with a precoding matrix $\mathbf{P}\in\mathbb{C}^{N_{\mathrm{tx}}\times N_{\mathrm{layer}}}$ to form the transmit signal $\mathbf{x}$. The selection of $N_{\mathrm{layer}}$ and corresponding $\mathbf{P}$ can be dynamically adapted to channel conditions. To mitigate CSI feedback overhead, the whole band of $N_{\mathrm{sc}}$ subcarriers is uniformly divided into $N_{\mathrm{sb}}$ subbands, with all symbols corresponding to the same subband sharing a common precoding matrix. The subband index and layer index are represented by $k\in\{1,\ldots,N_{\mathrm{sb}}\}$ and $n_{\mathrm{layer}}\in\{1,\ldots,N_{\mathrm{layer}}\}$, respectively. Subsequently, the precoded symbols are transformed into OFDM waveform through an inverse fast Fourier transform (IFFT) operation followed by cyclic prefix (CP) insertion.

At the UE side, the received time-domain signal is transformed into frequency domain via fast Fourier transform (FFT) after CP removal. Then the equivalent channel 
$\mathbf{H}_{\mathrm{eq}}\in\mathbb{C}^{N_{\mathrm{rx}}\times N_{\mathrm{layer}}}$ can be estimated for each RE based on the received DMRS. Frequency domain signal equalization is subsequently performed through linear minimum mean square error (LMMSE) filtering as, 
\begin{equation}\label{eq2}
\hat{\mathbf{x}}_{f,t} = \mathbf{H}_{\mathrm{eq}}^{\mathrm{H}} \left( \mathbf{H}_{\mathrm{eq}}\mathbf{H}_{\mathrm{eq}}^{\mathrm{H}} + \sigma_{n}^{2}\mathbf{I} \right)^{-1} \mathbf{y}_{f,t}
\end{equation}
where $\hat{\mathbf{x}}_{f,t}\in\mathbb{C}^{N_{\mathrm{layer}}\times 1}$ is the equalized signal and $\mathbf{I}$ is the identity matrix. Next, the equalized signal $\hat{\mathbf{x}}$ is fed into the demodulator, which calculates the log-likelihood ratios (LLRs). These LLRs are then input to the LDPC decoder to recover the original bits $\hat{\mathbf{b}}\in\{0, 1\}^{N_{\mathrm{b}}\times 1}$.

\subsection{CSI Feedback via Uplink Channel}
To facilitate channel-adaptive downlink precoding, the UE is required to calculate and report subband-level CSI to the BS. For simplicity, this article assumes perfect channel matrix $\mathbf{H}$ availability at the UE for CSI feedback and no feedback delay is considered. With regard to the $k$th subband, the spatial covariance matrix
$\mathbf{R}_{k}\in\mathbb{C}^{N_{\mathrm{tx}}\times N_{\mathrm{tx}}}$ is computed and averaged across its subcarriers as, 
\begin{equation}\label{eq3}
\mathbf{R}_{k}=\frac{1}{N_{\mathrm{t}}N_{\mathrm{sc}}/N_{\mathrm{sb}}} \sum_{t=1}^{N_{\mathrm{t}}} \sum_{f=(k-1)N_{\mathrm{sc}}/N_{\mathrm{sb}}+1}^{kN_{\mathrm{sc}}/N_{\mathrm{sb}}} \mathbf{H}_{f,t}^{\rm H} \mathbf{H}_{f,t}
\end{equation}
Then the full set of eigenvectors 
$\mathbf{v}_{k}\in\mathbb{C}^{N_{\mathrm{tx}}\times N_{\mathrm{tx}}}$ for $k$th subband can be calculated via eigen decomposition as,
\begin{equation}\label{eq4}
\mathbf{R}_{k} = \mathbf{v}_{k} \mathbf{\Lambda}_{k} \mathbf{v}_{k}^{\rm H}
\end{equation}
where
$\mathbf{\Lambda}_{k} = \mathrm{diag}\{ \lambda_{k,1}, \lambda_{k,2}, \ldots, \lambda_{k,N_{\mathrm{tx}}} \}$ is a diagonal matrix of eigenvalues sorted in descending order. The top $N_{\mathrm{layer}}$ eigenvectors,  concatenated and denoted as $\mathbf{w}_{k}\in\mathbb{C}^{N_{\mathrm{tx}}N_{\mathrm{layer}} \times 1}$, will be extracted, and therefore the CSI matrix to be fed back can be written as,
\begin{equation}\label{eq5}
\mathbf{W} = [\mathbf{w}_1, \mathbf{w}_2, \cdots, \mathbf{w}_{N_{\mathrm{sb}}}]^{\mathrm{T}} \in \mathbb{C}^{ N_{\mathrm{sb}} \times N_{\mathrm{tx}}N_{\mathrm{layer}}}
\end{equation}

With the preprocessed CSI available, the UE compresses it and feeds it back to the BS via the noisy uplink channel, enabling the BS to reconstruct the CSI for downlink precoding. In 5G systems, the standardized Type I and enhanced Type II (eType II) codebooks are widely used for CSI feedback \cite{TS0000,TS0001,TS0002,TS0003}, where the UE selects a precoding matrix from predefined codebooks and reports the precoding matrix index (PMI), allowing the BS to reconstruct the CSI accordingly. Additionally, AI/ML-based CSI feedback has been discussed in 3GPP, employing paired encoder-decoder NNs $f_{\rm e}(\cdot;\Theta_{\rm{E}})$ and $f_{\rm d}(\cdot;\Theta_{\rm{D}})$ with trainable parameters $\Theta_{\rm{E}}$ and $\Theta_{\rm{D}}$ at the UE and BS, respectively. Thus the AI/ML-based autoencoder $f_{\rm a}(\cdot; \Theta_{\rm{A}})$ with trainable parameters $\Theta = \{\Theta_{\rm{E}}, \Theta_{\rm{D}}\}$ for CSI feedback can be represented as
\begin{equation}\label{autoencoder_1}
\begin{split}
\mathbf{P} =  f_{\rm d}(n_{\rm{channel}}(f_{\rm e}(\mathbf{W};\Theta_{\rm{E}})); \Theta_{\rm{D}}) = f_{\rm a}(\mathbf{W};\Theta_{\rm{A}})
\end{split}
\end{equation}
where $f_{\rm e}(\cdot;\Theta_{\rm{E}})$ and $f_{\rm d}(\cdot;\Theta_{\rm{D}})$ implement the CSI compression and reconstruction processes respectively. Note that $n_{\rm{channel}}(\cdot)$ denotes the practical noisy uplink channel, where the decoder at BS tolerates the lossy input information and introduces disturbances to the decoder output, which can be comparable to and extended to the scenarios with disturbances caused by CSI errors introduced at the UE side.

\begin{figure*}[tbp]
 \begin{center}
 \includegraphics[width=1\textwidth]{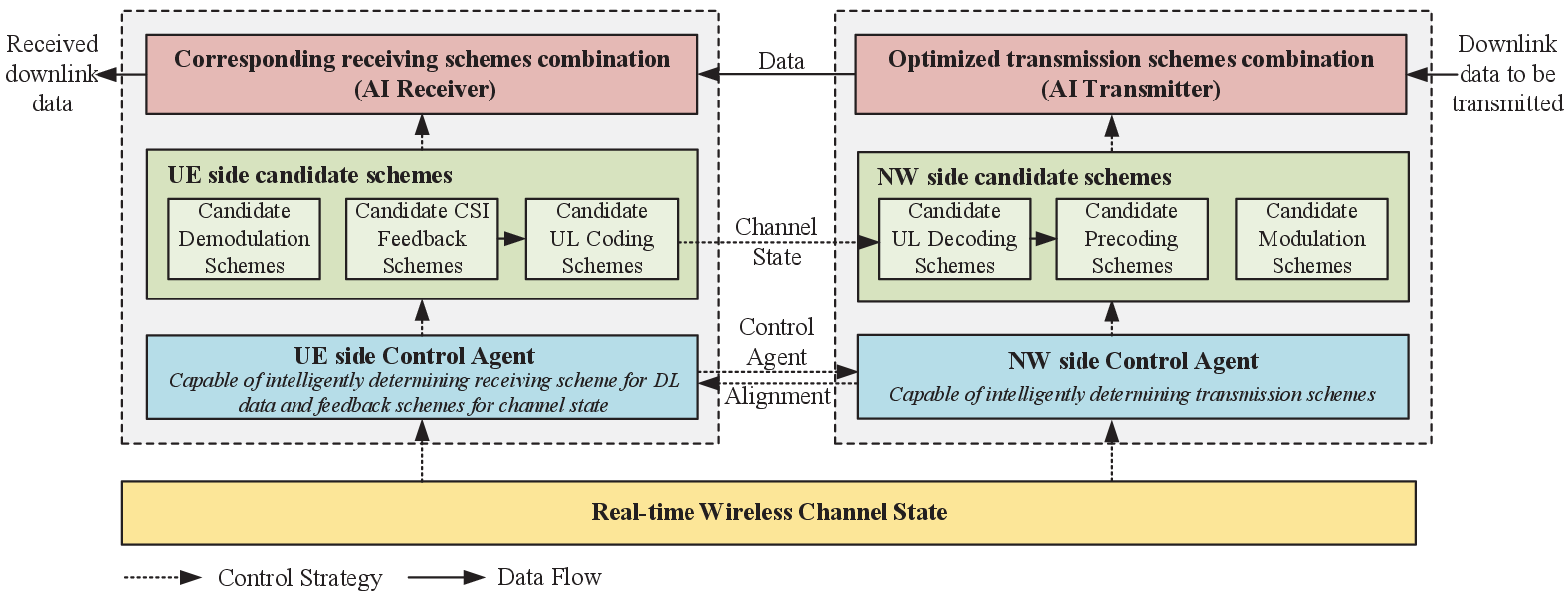}
 \end{center}
 \caption{Illustration of proposed AI-native framework }
 \label{figfr}
\end{figure*}

\section{AI-Native Cross-Module Optimized Physical Layer with Cooperative Control Agents}\label{sectionIII}
\subsection{Problem Formulation}\label{framework_sec}

As illustrated in the Fig. \ref{figfr}, the proposed AI-native framework includes the data flow and control strategy. For downlink data flow, it requires multiple candidate transmission schemes consisting of AI/ML features, e.g., AI/ML based modulation, AI/ML based precoding, AI/ML based CSI feedback and so on. Specifically, different compositions of the schemes constitute different physical layer links $\mathfrak{F}$, i.e.,
\begin{equation}
\label{F_1}
\begin{split}
\mathfrak{F} = \{\mathfrak{M}_q(\Theta_{\mathrm{mod}}, \Theta_{\mathrm{demod}}),\mathfrak{P}_v(\Theta_{\mathrm{P}}),\mathfrak{C}_z(\Theta_{\mathrm{E}}, \Theta_{\mathrm{D}}), \mathfrak{O}\}
\end{split}
\end{equation}
where $\mathfrak{M}_q(\cdot)$, $1 \leq q \leq Q$, $\mathfrak{P}_v(\cdot)$, $1 \leq v \leq V$ and $\mathfrak{C}_z(\cdot)$, $1 \leq z \leq Z$ denote the utilized modulation, precoding and CSI feedback schemes with number of candidate model architectures $Q$, $V$ and $Z$, respectively. Specifically, $\Theta_{\mathrm{mod}}$ and $\Theta_{\mathrm{demod}}$ represent the model parameters of the modulation and demodulation part of $\mathfrak{M}_q(\cdot)$, respectively. $\Theta_{\mathrm{P}}$ denotes the model parameters of $\mathfrak{P}_v(\cdot)$. $\Theta_{\mathrm{E}}$ and $\Theta_{\mathrm{D}}$ denote the model parameters of the encoder and decoder part of $\mathfrak{C}_z(\cdot)$, respectively. $\mathfrak{O}$ denotes the other features that are set as default in this paper. Moreover, in order to leverage the potential of these AI/ML features, targeted enhancement mechanisms for each one and cross-module optimization are also proposed, which will be described in detail later.

Since these AI/ML features leverage data-driven implementation to adapt to specific configurations or scenarios, they still require demand-aware control strategies to dynamically adjust transmission schemes according to real-time channel conditions. Therefore, for the control strategy design, this work proposes an AI/ML-based control agent $\mathfrak{G}(\cdot)$, which is designed to identify, manage, and coordinate various AI/ML features, particularly those implemented using two-sided models, i.e.,
\begin{equation}\label{control_agent_A}
\begin{split}
q,v,z = \mathfrak{G}(I;\Phi)
\end{split}
\end{equation}
where $I$ denotes the information of real-time channel conditions and $\Phi$ denotes the model parameters of $\mathfrak{G}$. When actually deploying, the architecture requires alignment between UE and NW side control agents through online or offline mechanisms first. Subsequently, the matched control agents on both sides can select and combine adaptable transmission schemes according to real-time channel conditions, while ensuring end-to-end compatibility. This strategy may significantly reduce the dedicated control signaling and procedures previously required for individual AI/ML feature alignment, enabling implicit configuration through native AI/ML-based control agents.

By combining the above data flow (\ref{F_1}) and control strategy (\ref{control_agent_A}), we formulate the problem as
\begin{equation}\label{frameworkeq}
\begin{aligned}
&\max_{\Theta, \mathfrak{F}} \frac{1}{N_{\mathrm{sc}}N_{\mathrm{t}}}p(\mathbf{b}, \hat{\mathbf{b}}) \\
&\ \mathrm{s.t.} \quad \Theta = \{\Theta_{\mathrm{mod}}, \Theta_{\mathrm{demod}}, \Theta_{\mathrm{P}},\Theta_{\mathrm{E}}, \Theta_{\mathrm{D}}, \Phi\} \\
&\qquad \ \  \mathfrak{F} = \{\mathfrak{M}_q(\Theta_{\mathrm{mod}}, \Theta_{\mathrm{demod}}),\mathfrak{P}_v(\Theta_{\mathrm{P}}),\mathfrak{C}_z(\Theta_{\mathrm{E}}, \Theta_{\mathrm{D}}), \mathfrak{O}\}\\
&\qquad \ \  q,v,z = \mathfrak{G}(I;\Phi)
\end{aligned}
\end{equation}
where the objective function maximizes the spectral efficiency. $\mathbf{b}\in\{0, 1\}^{N_{\mathrm{b}}\times 1}$ and $\hat{\mathbf{b}}\in\{0, 1\}^{N_{\mathrm{b}}\times 1}$ denote the original and received information bit stream through the data flow of link $\mathfrak{F}$, respectively. $p(\cdot)$ outputs the number of bits transmitted correctly. Since the above objective involves the design and optimization of multiple parts, it is difficult to solve in one step. Therefore, we consider breaking down the above objective and solving it through a series of enhancement mechanisms.

\subsection{Framework Construction}

\begin{figure*}[tbp]
 \begin{center}
 \includegraphics[width=1\textwidth]{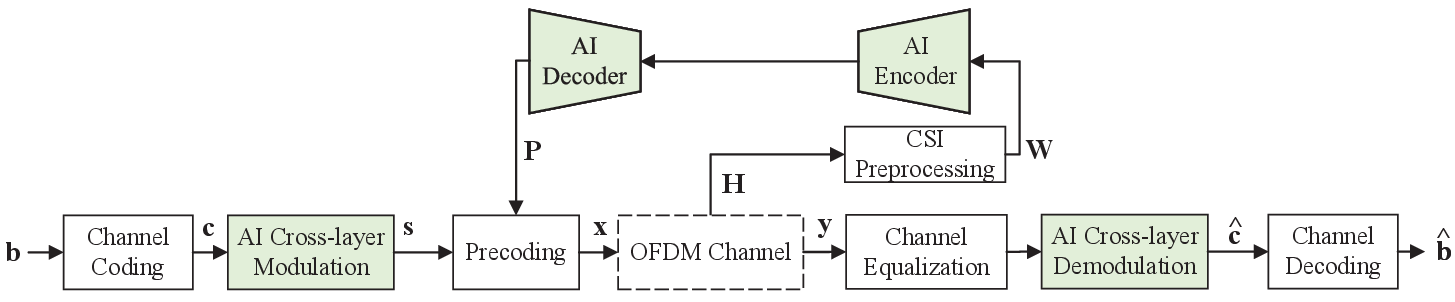}
 \end{center}
 \caption{Illustration of proposed AI-native cross-module optimized transmission scheme. }
 \label{fig0}
\end{figure*}

Here we first disassemble the proposed AI-native physical layer link $\mathfrak{F}$ in more detail. As illustrated in Fig. \ref{fig0}, it employs a hybrid design approach that seamlessly integrates AI/ML components with physical layer while maintaining structural compatibility with standardized interfaces. Specifically, the transmitter consists of channel encoding, modulation, precoding and OFDM mapping, while the receiver consists of channel estimation, equalization, demodulation and channel decoding, where the AI/ML models are strategically embedded to enhance multiple critical functions of source coding, channel coding, modulation, precoding and CSI feedback.

Unlike conventional systems where individual components are optimized in isolation, this architecture enables coordinated learning across the end-to-end link through a differentiable processing pipeline encompassing all modules. The joint training mechanism coordinates AI/ML modules to not only excel in their specific functions but also facilitate mutual adaptation among AI/ML components and the link, where the proposed enhancements expand the solution space for enhancing the modules and potentially achieving improved performance, which will be described in the following parts.

\subsubsection{Constellation Solution Space Expansion for Downlink Transmission}\label{modluationCrosslayer1111}

In this subsection, the enhancement for modulation scheme $\mathfrak{M}$ is introduced, where an AI/ML-based cross-layer modulation scheme is proposed for downlink transmission. Conventional modulation schemes in existing systems perform independent constellation mapping in isolated two-dimensional complex planes per transmission layer, i.e., 
\begin{equation}\label{eqmod1}
s_{f,t,n_{layer}} = \mathrm{mod}(\mathbf{c}_{f,t,n_{\mathrm{layer}}}), n_{\mathrm{layer}}=1,\dots,N_{\mathrm{layer}}
\end{equation}
where $s_{f,t,n_{\mathrm{layer}}}\in\widetilde{\mathbb{C}}$ and $\mathbf{c}_{f,t,n_{\mathrm{layer}}}\in\{0, 1\}^{m\times 1}$ denote the modulated symbol and coded bits of subcarrier $f$, OFDM symbol $t$ and layer $n_{\mathrm{layer}}$, respectively. $\widetilde{\mathbb{C}}$ denotes the constellation set and $\mathrm{mod}(\cdot)$ denotes the traditional modulation mapping. It can be noted that the independent modulation design across layers inherently limits the solution space for optimal constellation exploration, leaving significant room for improving modulation performance.

To address the aforementioned challenges and further expand the solution space of modulation constellation for multi-layer transmission and leverage the advantages of AI/ML, here we propose an approach that constructs a unified high-dimensional signal space integrating multiple layers, i.e.,
\begin{equation}\label{eqmod1b}
\tilde{\mathbf{s}}_{f,t} = g_{\mathrm{mod}}([\mathbf{c}_{f,t,1}^\mathrm{T},\dots,\mathbf{c}_{f,t,N_{\mathrm{layer}}}^\mathrm{T}]^\mathrm{T}, \Theta_{\mathrm{mod}})
\end{equation}
where $g_{\mathrm{mod}}(\cdot)$ represents the proposed AI/ML-based cross-layer modulator with trainable model weights of $\Theta_{\mathrm{mod}}$ and $\tilde{\mathbf{s}}_{f,t}\in\mathbb{C}^{N_{\mathrm{layer}}\times 1}$ denotes the modulated symbols vector of subcarrier $f$ and OFDM symbol $t$. Mirroring the AI/ML-based modulator, the demodulator treats multi-layer signals as a joint high-dimensional unit and demodulates the equalized signal to the LLRs, i.e.,
\begin{equation}\label{eqmod2}
[\hat{\mathbf{c}}_{f,t,1}^\mathrm{T},\dots,\hat{\mathbf{c}}_{f,t,N_{\mathrm{layer}}}^\mathrm{T}]^\mathrm{T} = g_{\mathrm{demod}}(\tilde{\mathbf{s}}^\mathrm{eq}_{f,t}, \Theta_{\mathrm{demod}})
\end{equation}
where $g_{\mathrm{demod}}(\cdot)$ represents the proposed AI/ML-based cross-layer demodulator with trainable model weights of $\Theta_{\mathrm{demod}}$, $\tilde{\mathbf{s}}^\mathrm{eq}_{f,t}\in\mathbb{C}^{N_{\mathrm{layer}}\times 1}$ and $\hat{\mathbf{c}}_{f,t,n_{\mathrm{layer}}}\in\{0, 1\}^{m\times 1}$ with $n_{\mathrm{layer}}=1,\dots,N_{\mathrm{layer}}$ denote the equalized signal and received LLRs of subcarrier $f$ and OFDM symbol $t$, respectively. Within the proposed method, $g_{\mathrm{mod}}(\cdot)$ and $g_{\mathrm{demod}}(\cdot)$ directly learn optimal bits-to-symbols mapping, breaking the dimensional isolation between layers in conventional methods. This approach enables joint exploitation of cross-layer features and expands the solution space for optimal constellation search, thereby facilitating the learning of a better constellation set to further enhance modulation performance.

\subsubsection{Utility-Oriented Precoder Construction}\label{Utility-Oriented}
In this subsection, the enhancement for precoding scheme $\mathfrak{P}$ is introduced, which enables a utility-oriented precoder construction by optimizing precoding performance from a comprehensive end-to-end perspective rather than confining the optimization scope solely to the precoding module itself. Reviewing common implementations of AI/ML-based CSI feedback in (\ref{autoencoder_1}), the optimization objective for AI/ML-based CSI feedback can generally be given as
\begin{equation}\label{score_function_1}
\begin{split}
\min_{\Theta_{\mathrm{A}}} -\rho(\mathbf{W}, \mathbf{P})
 = \min_{\Theta_{\mathrm{A}}} -\frac{1}{N_\textrm{sb}}\sum_{l=1}^{N_\textrm{sb}}\left(\frac{\|\mathbf{w}^{\rm H}\mathbf{p}\|_2}{\|\mathbf{w}\|_2\|\mathbf{p}\|_2}\right)^2
\end{split}
\end{equation}
where $\rho(\cdot) \in [0,1]$ denotes the squared generalized cosine similarity (SGCS) and a larger $\rho$ indicates higher CSI recovery accuracy. Here, $\|\cdot\|_2$ denotes the $\ell_{2}$ norm, $\mathbf{w}_{l}$ and $\mathbf{p}_{l}$ represent the original and recovered CSI eigenvector on the $l$-th subband, respectively. It can be noticed that the recovered CSI $\mathbf{P}$ is expected to be as close as possible to the original one $\mathbf{W}$ which is obtained by eigen decomposition of (\ref{eq4}). This ensures that $\mathbf{P}$ can be directly used as a precoder to preserve the spatial characteristics of the channel (e.g., the main eigendirections of $\mathbf{H}$), thereby enforcing layer-wise orthogonality to minimize inter-layer interference under the assumption of layer-independent modulation schemes. However, with the introduction of joint design of AI/ML-based modules, the precoder obtained by eigen decomposition is no longer optimal for the link with greater degrees of freedom, since optimization objective (\ref{score_function_1}) contributes indirectly to the final throughput performance, calling for a more appropriate precoder construction method.

To address the above issues, a utility-oriented precoder construction method is proposed, where the decoder function is redefined by shifting its role from recovering the original CSI to directly generating precoding matrices optimized for end-to-end performance, using binary cross-entropy error (BCE) loss function of

\begin{equation}
\begin{split}
&\min_{\hat{\Theta}_{\textrm{E}},\hat{\Theta}_{\textrm{D}}} \mathcal{L}_{\textrm{bce}}(\mathbf{c},\hat{\mathbf{c}})\\
=&\min_{\hat{\Theta}_{\textrm{E}},\hat{\Theta}_{\textrm{D}}}-\frac{1}{N_{\mathrm{sc}}N_{\mathrm{t}}N_{\mathrm{layer}}}\Bigg(\sum_{f=1}^{N_{\mathrm{sc}}}\sum_{t=1}^{N_{\mathrm{t}}}\sum_{n_{\mathrm{layer}}=1}^{N_{\mathrm{layer}}} c_{f,t,n_{\mathrm{layer}}}\\ &\cdot \log(\tilde{c}_{f,t,n_{\mathrm{layer}}}) + (1-c_{f,t,n_{\mathrm{layer}}})\log\left(1-\tilde{c}_{f,t,n_{\mathrm{layer}}}\right)\Bigg)
\end{split}
\end{equation}
where $\hat{\Theta}_{\rm{E}}$ and $\hat{\Theta}_{\rm{D}}$ denote the trainable model weights of proposed encoder and decoder respectively, and
\begin{equation}
\label{llr_ori}
\begin{split}
\tilde{c}_{f,t,n_{\mathrm{layer}}} = \frac{1}{1+e^{-\hat{c}_{f,t,n_{\mathrm{layer}}}}}
\end{split}
\end{equation}
denotes the transformation of LLR $\hat{c}_{f,t,n_{\mathrm{layer}}}$ through sigmoid function. On the one hand, this relaxation of constraints endows the model with greater solution space, relaxes rigid eigenstructure constraints of (\ref{score_function_1}) to construct precoding matrix tailored to cross-layer modulation, rather than being limited by the input CSI structure. On the other hand, the proposed objective optimizes the parameters of the encoder and decoder $\hat{\Theta}_{\textrm{E}}$ and $\hat{\Theta}_{\textrm{D}}$ respectively by aiming to maximize bit recovery accuracy so that realizes the solution of $\Theta_{\mathrm{E}}$ and $\Theta_{\mathrm{D}}$ in $\Theta$ in (\ref{frameworkeq}). To circumvent the undesired amplification of average signal power caused by interdependent coupling effects in the jointly optimized modulation and precoding schemes, a power normalization process is introduced for the precoded signal $\mathbf{x}$ before transmission, as described below,
\begin{equation}
x_{f,t,n_{\mathrm{tx}}} = \frac{x_{f,t,n_{\mathrm{tx}}}}{\sqrt{\frac{1}{N_{\mathrm{sc}}N_{\mathrm{t}}}\big(\sum_{f=1}^{N_{\mathrm{sc}}}\sum_{t=1}^{N_{\mathrm{t}}}\sum_{n_{\mathrm{tx}}=1}^{N_{\mathrm{tx}}}\|x_{f,t,n_{\mathrm{tx}}}\|_2\big)}}
\end{equation}
where $x_{f,t,n_{\mathrm{tx}}}\in\mathbb{C}$ with $n_{\mathrm{tx}}=1,\dots,N_{\mathrm{tx}}$ denote the precoded signal of subcarrier $f$, OFDM symbol $t$ and transmit antenna $n_{\mathrm{tx}}$, respectively.

\begin{table*}[b]
\centering
\caption{Model Structure Implementing Proposed Cross-layer Modulation}
\label{modNN}
\setlength{\tabcolsep}{4.4mm}{
\begin{tabular}{|c|c|c|c|}
\hline
 & Layer & Parameters & Output dimension   \\ \hline
\multirow{4}{*}{Modulator $g_{\mathrm{mod}}(\cdot)$}         & Input & / & $N_{\mathrm{batch}}\times N_{\mathrm{sc}}N_{\mathrm{t}} \times mN_{\mathrm{layer}}$  \\ \cline{2-4}
& Dense $\times N_{\mathrm{dense}}$ & \makecell{Units=$K_{\mathrm{dense}}$,  Activation=ReLU} & $N_{\mathrm{batch}}\times N_{\mathrm{sc}}N_{\mathrm{t}} \times K_{\mathrm{dense}}$  \\ \cline{2-4}
& Dense  & Units=$N_{\mathrm{RI}}N_{\mathrm{layer}}$ & $N_{\mathrm{batch}}\times N_{\mathrm{sc}}N_{\mathrm{t}} \times N_{\mathrm{RI}}N_{\mathrm{layer}}$  \\ \cline{2-4}
& Normalization  & / & $N_{\mathrm{batch}}\times N_{\mathrm{sc}}N_{\mathrm{t}} \times N_{\mathrm{RI}}N_{\mathrm{layer}}$  \\ \hline
\multirow{4}{*}{Demodulator $g_{\mathrm{demod}}(\cdot)$}         & Input & / & $N_{\mathrm{batch}}\times N_{\mathrm{rx}} \times N_{\mathrm{sc}}N_{\mathrm{t}} \times N_{\mathrm{RI}}N_{\mathrm{layer}}$  \\ \cline{2-4}
& Conv2D & Kernel=$1\times1$, Filters=$D$  & $N_{\mathrm{batch}}\times N_{\mathrm{sc}}N_{\mathrm{t}} \times D$  \\ \cline{2-4}
& Residual Block $\times N_{\mathrm{res}}$ & Kernel=$1\times1$, Filters=$D$  & $N_{\mathrm{batch}}\times N_{\mathrm{sc}}N_{\mathrm{t}} \times D$  \\ \cline{2-4}
& Conv2D & Kernel=$1\times1$, Filters=$mN_{\mathrm{layer}}$  & $N_{\mathrm{batch}}\times N_{\mathrm{sc}}N_{\mathrm{t}} \times mN_{\mathrm{layer}}$  \\ \hline
\end{tabular}}
\end{table*}

\begin{algorithm}[t]
\caption{Cross-Module Optimization Process}
\label{alg_1}
\textbf{Initialization}: Randomize $\hat{\Theta}_{\textrm{E}}$, $\hat{\Theta}_{\textrm{D}}$, $\hat{\Theta}_{\textrm{mod}}$, $\hat{\Theta}_{\textrm{demod}}$, $\Phi$; $\lambda =0.5$;\\ 
\textbf{Phase 1: pre-convergence for cross-layer modulation}\\
\ \ \ Update $\hat{\Theta}_{\textrm{E}}$, $\hat{\Theta}_{\textrm{D}}$, $\hat{\Theta}_{\textrm{mod}}$ and $\hat{\Theta}_{\textrm{demod}}$ by solving (\ref{optiAAAA});\\
\textbf{Phase 2: objective relaxation for utility-oriented precoding}\\
\ \ \ Update $\hat{\Theta}_{\textrm{E}}$, $\hat{\Theta}_{\textrm{D}}$, $\hat{\Theta}_{\textrm{mod}}$ and $\hat{\Theta}_{\textrm{demod}}$ by solving (\ref{optiAAAA}) with $\lambda = 1$;\\
\textbf{Phase 3: strategy learning of control agents}\\
\ \ \ Update $\Phi$ by solving (\ref{agentoptiii}).
\end{algorithm}

\subsubsection{Bit-Level Bottlenecks Bypass for Uplink CSI Feedback}\label{bitbypass}
In this subsection, the enhancement for CSI feedback scheme $\mathfrak{C}$ is introduced. For existing AI/ML-based CSI feedback, encoder compresses and quantizes the original CSI $\mathbf{W}$ to a bitstream $\mathbf{b}_\mathrm{CSI} \in \{0,1\}^{N_\mathrm{CSIraw}\times 1} $, from which $\mathbf{W}$ can be recovered by decoder. The channel coding can also be jointly implemented by the encoder and decoder, directly processing the CSI into the channel-coded bitstream $\mathbf{c}_\mathrm{CSI} \in \{0,1\}^{N_\mathrm{CSIcoded}\times 1}$ where the coding rate is calculated as $N_\mathrm{CSIraw} / N_\mathrm{CSIcoded}$. However, since the intermediate variables between the encoder and decoder are in bit form, it brings a series of bit-level bottlenecks. First, it introduces unavoidable quantization errors. Second, the non-differentiable gradient caused by quantization increases the difficulty of training, thereby limiting the performance. Third, traditional modulation for bit transmission also brings limitation to the solution space for optimizing modulation constellation.

To address the above challenges, we propose to incorporate modulation and demodulation into the encoder $\hat{f}_{\rm e}(\cdot)$ and decoder $\hat{f}_{\rm d}(\cdot)$ respectively, i.e.,
\begin{equation}
\begin{split}
\mathbf{s}_\mathrm{CSI} =  \hat{f}_{\rm e}(\mathbf{W};\hat{\Theta}_{\rm{E}})
\end{split}
\text{,}
\end{equation}
\begin{equation}
\begin{split}
\mathbf{P} =  \hat{f}_{\rm d}(\hat{\mathbf{s}}_\mathrm{CSI};\hat{\Theta}_{\rm{D}})
\end{split}
\end{equation}
where $\hat{\Theta}_{\rm{E}}$ and $\hat{\Theta}_{\rm{D}}$ denote the trainable model weights of encoder and decoder respectively, $\mathbf{s}_\mathrm{CSI} \in \mathbb{C}^{N_{\mathrm{CSI}}\times 1}$ denotes the symbols for transmission in MIMO-OFDM system, $\hat{\mathbf{s}}\in \mathbb{C}^{N_{\mathrm{CSI}}\times 1}$ denotes the received symbols after channel equalization, and $N_{\mathrm{CSI}} = N_\mathrm{CSIcoded} / m$ with $m$ denoting the modulation order. Notably, the transition of the autoencoder’s intermediate variable from bit-form $\mathbf{c}_\mathrm{CSI}$ to non-bit-form $\mathbf{s}_\mathrm{CSI}$ eliminates quantization errors and training difficulties, where the continuous solution space permits gradient-based training to converge to local optima without being trapped by quantization boundaries, and the preserved differentiability throughout the system allows joint training of constellation geometry with other modules, aligning the design with global performance objectives rather than isolated CSI reconstruction metrics. While absolute optimality cannot be guaranteed due to the non-convex nature of the problem, this expanded solution space provides necessary conditions for finding superior solutions, as empirically verified in the simulation results of subsequent Section \ref{SectionIV}.

\subsubsection{Control Agent for Optimized Transmission Schemes}\label{controlAgentttt}
In this subsection, the design for control agent $\mathfrak{G}$ is introduced. From the perspective of system gain, the traditional link adaptation mechanism utilizes a lookup table mapping CQI to MCS, which is based on the BLER threshold analyzed from traditional non-AI-native links. However, once AI/ML features are introduced, the mapping rules are no longer easy to obtain by theoretical derivation due to the uninterpretability of AI/ML models. At the same time, the implementation form of link adaptation also changes from MCS selection to model switching. Existing solutions of LCM \cite{chen20235g} adopt post-hoc adaptation, as the switch is conducted after performance degradation is detected during model monitoring. These call for novel design of control strategy for the optimized transmission scheme.

Here we propose an AI/ML-based strategy that directly learns the transmission schemes selection using the data collected from the proposed AI-native link and target channel scenario. First, there is a one-to-one correspondence between the utility-oriented precoder and the CSI input to the encoder, where the utility-oriented precoder and the cross-layer-modulated symbols can adjust the transmission power of each layer, and indirectly adjust the number of transmission layers, e.g., the layers with zero power are not sent. Therefore, the proposed CSI feedback not only directly adjusts the precoder, but also indirectly adjusts the number of transmission layers. Second, AI/ML-based control agents are also proposed that learn the mapping between estimated channel quality of signal to interference and noise ratio (SINR), and the transmission schemes that optimize the throughput of AI-native link, i.e.,
\begin{equation}
\begin{split}
\mathbf{v} =  g_{\mathrm{la}}(\mathbf{q};\Phi)
\end{split}
\end{equation}
where $g_{\mathrm{la}}(\cdot)$ denotes the switching model with trainable model weights $\Phi$, $\mathbf{q} \in \mathbb{C}^{N_{\mathrm{layer}} \times 1}$ denotes the estimated SINR of $N_{\mathrm{layer}}$ layers, and $\mathbf{v} \in [0,1]^{J \times 1}$ selects the optimal one from $J$ potential models, and 1 means the one is selected, and vice versa.

\begin{table*}[t]
\centering
\caption{Model Structure Implementing Proposed CSI Feedback}
\label{csiNN}
\setlength{\tabcolsep}{3.8mm}{
\begin{tabular}{|c|c|c|c|}
\hline
 & Layer & Parameters & Output dimension   \\ \hline
\multirow{7}{*}{Encoder $\hat{f}_{\mathrm{e}}(\cdot)$}         & Input & / & $N_{\mathrm{batch}} \times  N_{\mathrm{sb}} \times N_{\mathrm{tx}} N_{\mathrm{layer}}$  \\ \cline{2-4}
& $\mathbb{C}2\mathbb{R}$ \& Reshape  & Shape=$N_{\mathrm{batch}} \times N_{\mathrm{sb}} \times N_{\mathrm{RI}}N_{\mathrm{tx}}N_{\mathrm{layer}}$ & $N_{\mathrm{batch}} \times N_{\mathrm{sb}} \times N_{\mathrm{RI}}N_{\mathrm{tx}}N_{\mathrm{layer}}$  \\ \cline{2-4}
& Dense  & Units=$N_{\mathrm{embedding}}$ & $N_{\mathrm{batch}} \times N_{\mathrm{sb}} \times N_{\mathrm{embedding}}$  \\ \cline{2-4}
& Transformer Block $\times N_\mathrm{transformer}$   & Head=$N_{\mathrm{head}}$, Embedding=$N_{\mathrm{embedding}}$ & $N_{\mathrm{batch}} \times N_{\mathrm{sb}} \times N_{\mathrm{embedding}}$  \\ \cline{2-4}
& LayerNormalization \& Dense   & Units=$N_{\mathrm{RI}}N_{\mathrm{tx}}N_{\mathrm{layer}}$ & $N_{\mathrm{batch}} \times N_{\mathrm{sb}} \times N_{\mathrm{RI}}N_{\mathrm{tx}}N_{\mathrm{layer}}$  \\ \cline{2-4}
& Reshape  & Shape=$N_{\mathrm{batch}} \times N_{\mathrm{sb}}N_{\mathrm{RI}}N_{\mathrm{tx}}N_{\mathrm{layer}}$ & $N_{\mathrm{batch}} \times N_{\mathrm{sb}}N_{\mathrm{RI}}N_{\mathrm{tx}}N_{\mathrm{layer}}$  \\ \cline{2-4}
& Dense \& Normalization  & Units=$N_{\mathrm{RI}}N_{\mathrm{CSI}}$ & $N_{\mathrm{batch}} \times N_{\mathrm{RI}}N_{\mathrm{CSI}}$  \\ \hline

\multirow{7}{*}{Decoder $\hat{f}_{\mathrm{d}}(\cdot)$}         & Input & / & $N_{\mathrm{batch}} \times N_{\mathrm{RI}}N_{\mathrm{CSI}}$   \\ \cline{2-4}
& Dense  & Units=$N_{\mathrm{sb}}N_{\mathrm{RI}}N_{\mathrm{tx}}N_{\mathrm{layer}}$ & $N_{\mathrm{batch}} \times N_{\mathrm{sb}}N_{\mathrm{RI}}N_{\mathrm{tx}}N_{\mathrm{layer}}$  \\ \cline{2-4}
& Reshape  & Shape=$N_{\mathrm{batch}} \times N_{\mathrm{sb}}\times N_{\mathrm{RI}}N_{\mathrm{tx}}N_{\mathrm{layer}}$ & $N_{\mathrm{batch}} \times N_{\mathrm{sb}} \times N_{\mathrm{RI}}N_{\mathrm{tx}}N_{\mathrm{layer}}$  \\ \cline{2-4}
& Dense  & Units=$N_{\mathrm{embedding}}$ & $N_{\mathrm{batch}}\times N_{\mathrm{sb}} \times N_{\mathrm{embedding}}$  \\ \cline{2-4}
& Transformer Block $\times N_\mathrm{transformer}$   & Head=$N_{\mathrm{head}}$, Embedding=$N_{\mathrm{embedding}}$ & $N_{\mathrm{batch}} \times N_{\mathrm{sb}} \times N_{\mathrm{embedding}}$  \\ \cline{2-4}
& LayerNormalization \& Dense   & Units=$N_{\mathrm{RI}}N_{\mathrm{tx}}N_{\mathrm{layer}}$ & $N_{\mathrm{batch}} \times N_{\mathrm{sb}} \times N_{\mathrm{RI}}N_{\mathrm{tx}}N_{\mathrm{layer}}$  \\ \cline{2-4}
& $\mathbb{R}2\mathbb{C}$ \& Reshape  & Shape=$N_{\mathrm{batch}} \times N_{\mathrm{sb}} \times N_{\mathrm{tx}}N_{\mathrm{layer}}$ & $N_{\mathrm{batch}} \times N_{\mathrm{sb}} \times N_{\mathrm{tx}} N_{\mathrm{layer}}$  \\ \hline
\end{tabular}}
\end{table*}

\subsubsection{Cross-Module Optimization Process}
In this subsection, the building process for link $\mathfrak{F}$ with agent $\mathfrak{G}$ is introduced. As the solution space of the overall link expands with the introduction of control agents, model training becomes more difficult than training a single AI/ML feature, so a targeted training strategy needs to be carefully designed. The learning strategy employs three progressively refined training phases to balance convergence stability with optimization freedom. Specifically, phase I serves as a pre-training stage where the cross-layer modulation models $\hat{g}_{\rm mod}(\cdot)$ and $\hat{g}_{\rm demod}(\cdot)$, and CSI feedback models $\hat{f}_{\rm e}(\cdot)$ and $\hat{f}_{\rm d}(\cdot)$ are trained jointly, reaching a preliminary convergence, i.e.,
\begin{equation}
\label{optiAAAA}
\begin{split}
\min_{\hat{\Theta}_{\textrm{E}},\hat{\Theta}_{\textrm{D}},\hat{\Theta}_{\textrm{mod}},\hat{\Theta}_{\textrm{demod}}}\lambda\mathcal{L}_{\textrm{bce}}(\mathbf{c},\hat{\mathbf{c}}) -(1-\lambda)\rho(\mathbf{W}, \mathbf{P})
\end{split}
\end{equation}
where $\lambda=0.5$ and it essentially builds on the initialization point of the models where cross-layer modulation accommodates a traditional eigen-decomposition-based precoding, ensuring stable preliminary convergence. 

Phase II then deactivates the above auxiliary loss with $\lambda=1$, focusing exclusively on BCE-driven end-to-end refinement to finally achieve the proposed cross-layer modulation and utility-oriented precoder. Notably, channel encoding and decoding modules are excluded during the training phase to ease the difficulty of training and are cascaded only in the inference phase to maintain compatibility with standard communication pipelines. It should be noted that the proposed bit-level bottlenecks bypass solution maintains the complete link differentiable for the above gradient-based optimization. 

As for Phase III, the switching model $g_{\mathrm{la}}(\cdot)$ is trained using the loss of categorical cross-entropy, i.e.,
\begin{equation}
\label{agentoptiii}
\begin{split}
\min_{\Phi} \mathcal{L}_{\textrm{cce}}(\mathbf{v},\hat{\mathbf{v}}) =\min_{\Phi}-\frac{1}{J}\sum_{j=1}^{J} v_{j}\log(\hat{v}_{j})
\end{split}
\end{equation}
where $\mathbf{v} = [v_1,\dots,v_J]$ denotes the label indicating the selection of optimal models that maximize spectral efficiency in (\ref{frameworkeq}), and $\hat{\mathbf{v}} = [\hat{v}_1,\dots,\hat{v}_J]$ denotes the temporary output of $g_{\mathrm{la}}(\cdot)$ during training.

The proposed cross-module optimization process can be summarized in Algorithm \ref{alg_1}. Here, each training step updates parameters using gradient descent based on the average loss over $N_{\mathrm batch}$ independent communication trials. For each communication trial, the transmitted bits are randomly generated, and the channel model parameters are stochastically initialized to simulate diverse propagation conditions. Furthermore, the signal-to-noise ratio (SNR) is uniformly randomized within a predefined operational range per communication trial, ensuring robustness across varying noise conditions. We also recognize that training complexity increases with system dimensions.

\subsubsection{AI/ML Model Structure Implementation}
The model structure supporting the proposed AI/ML-based cross-layer modulation in \ref{modluationCrosslayer1111} is detailed in Table \ref{modNN}. For modulator $g_{\mathrm{mod}}(\cdot)$, $N_{\mathrm{dense}} = 4$ fully connected (Dense) layers with units of $K_{\mathrm{dense}}=256$ and rectified linear unit (ReLU) activation are utilized for feature extraction. While the output layer is also implemented using one Dense layer with units of $N_{\mathrm{RI}}\times N_{\mathrm{layer}}$, where $N_{\mathrm{RI}} = 2$ denotes the separated real and imaginary parts. Finally, the output symbols are divided by the average energy to ensure normalized constellation. For demodulator $g_{\mathrm{demod}}(\cdot)$. The well-known ResNet-inspired demodulator employs an input two-dimensional convolutional layer (Conv2D), $N_{\mathrm{res}}=4$ residual blocks and an output Conv2D with filters $D=256$ to transform equalized signals into LLRs, where double sequential batch normalization, ReLU activation and Conv2D with residual connection are implemented in each residual block. Note that all convolutional layers employ $1 \times 1$ kernel size to enable independent processing of equalized signals per RE.

The model structure supporting the proposed CSI feedback in \ref{Utility-Oriented} and \ref{bitbypass} is detailed in Table \ref{csiNN}. The Transformer backbone for CSI feedback namely EVCsiNet-T \cite{han2024ai} with embedding dimension of $N_{\mathrm{embedding}}=256$, $N_{\mathrm{head}}=4$ heads and $N_{\mathrm{transformer}} = 6$ basic blocks is implemented for feature extraction, wherein the EVCsiNet-T is a common model structure used by 3GPP to evaluate the performance of CSI feedback. Specifically, the quantization and dequantization layers are replaced by Dense layers with normalization that directly output normalized modulation constellation symbols.

As for the proposed control agents in \ref{controlAgentttt}, two Dense layers sandwiched by a batch normalization are utilized to implement $g_{\mathrm{la}}(\cdot)$, with units of $4N_{\mathrm{layer}}$ and $N_{\mathrm{layer}}$, and activations of ReLU and Softmax, respectively.

{\color{black}
\subsection{Theoretical analysis}
In this subsection, we take the module of multi-layer modulation as an instance, shedding light on the potential benefits for performing cross-layer modulation with theoretical analysis.
In our proposed transmission scheme, the bit-interleaved coded modulation (BICM) design principle is still inherited, where the performance of modulation can be measured by BICM capacity \cite{i2008bit}
\begin{equation}
\label{bicm_capacity}
\begin{split}
C=m-\sum_{i=1}^m \mathbbm{E} \left[ {\rm{log_2}} \frac{\sum_{{\bf{x}}\in \mathcal{X}} p({\bf{y}}|{\bf{x}})}{\sum_{{\bf{x}} \in \mathcal{X}_b^i} p({\bf{y}}|{\bf{x}})} \right], 
\end{split}
\end{equation}
where $b \in \{0,1\}$ is equiprobable bit, $m$ denotes modulation order, ${\bf{y}}, {\bf{x}} \in {\mathbbm{C}}^{N_{\rm{layer}}}$ denote the received constellation and transmitted constellation, respectively. Notation $\mathcal{X}_b^i$ represents the set of symbols where the $i$-th bit equals $b$ and $\mathcal{X}$ denotes the set of all symbols. For traditional modulation method, both ${\bf{y}}$ and ${\bf{x}}$ are complex scalars. On the contrary, ${\bf{y}}$ and ${\bf{x}}$ can be complex vectors. The expectation operator shall jointly consider the high-dimensional distribution of transmitted symbols and noise, which is hard to be directly analyzed.

In practice, the BICM capacity in (\ref{bicm_capacity}) could be approximated following the Monte-Carlo techniques, deriving
\begin{equation}
\label{bicm_capacity2}
\begin{aligned}
C&=m-\sum_{i=1}^m \mathbbm{E} \left[ {\rm{log_2}} \left( 1+ \frac{\sum_{{\bf{x}}\in \mathcal{X}_{b'}^i} p({\bf{y}}|{\bf{x}})}{\sum_{{\bf{x}} \in \mathcal{X}_b^i} p({\bf{y}}|{\bf{x}})} \right) \right] \\
&\approx m-\frac{1}{S} \sum_{s=1}^S \sum_{i=1}^m \sum_{b=0}^1 p_b \left[ {\rm{log_2}} \left( 1+ l_b^{(i)}({\bf{y}}_s) \right) \right]
\end{aligned}
\end{equation}
where $b'$ denotes the flipped bit of $b$, $p_b$ denotes the probability of bit $b$, and $S$ is the total number of samplings. Notation $l_b^{(i)}({\bf{y_s}})$ can be expressed by
\begin{equation}
\label{llr}
\begin{aligned}
l_b^{(i)}({\bf{y_s}}) \triangleq \frac{\sum_{{\bf{x}}\in \mathcal{X}_{b'}^i} p({\bf{y_s}}|{\bf{x}})}{\sum_{{\bf{x}} \in \mathcal{X}_b^i} p({\bf{y_s}}|{\bf{x}})}.
\end{aligned}
\end{equation}
By assuming AWGN channel, item $p({\bf{y_s}}|{\bf{x}})$ has the form of
\begin{equation}
\label{awgn_prob}
\begin{aligned}
p({\bf{y_s}}|{\bf{x}}) = \frac{1}{\left(\sqrt{2\pi}\sigma_n \right)^{2N}} \exp \left(-\frac{||{\bf{y}}_s-{\bf{x}}||^2}{2\sigma_n^2} \right),
\end{aligned}
\end{equation}
where $\sigma_n^2$ denotes the variance of noise in each orthogonal component and $N_{\rm{layer}}$ is written as $N$ for short. Afterwards, the maximization of channel capacity in (\ref{bicm_capacity2}) can be approximated by minimizing $l_b^{(i)}({\bf{y_s}})$ for $b \in \{0,1\}$. In other words, the optimization target is to minimize the error probability for each transmitted symbol.

Due to the intractability of the optimization on $l_b^{(i)}({\bf{y_s}})$ for all symbols, we turn to the analysis of upper bound of error probability for each symbol, which indirectly affects channel capacity. Reviewing the exponential component in (\ref{awgn_prob}), the minimization of probability is equivalent to maximizing the Euclidean distance between adjacent constellations, i.e. maximizing the minimum distance $d_{\rm{min}}$. Intuitively, a larger $d_{\rm{min}}$ for the given dimension can effectively reduce the error probability and contributes to improved channel capacity.

Within the proposed modulation scheme $\mathfrak{M}$, the constraint of uniform QAM constellation is relaxed, and an arbitrarily distributed constellation can be designed in the given $2N$-dimensional space. Generally, the constellations are designed following the normalized power constraint. By assuming a continuous distribution of constellations [45], the averaged power of constellation design $\mathfrak{M}$ can be written as 
\begin{equation}
\label{power}
\begin{aligned}
P_{\mathfrak{M}} = \mathbbm{E} ||{\bf{x}}||^2 \approx \int_{\mathbbm{C}^{N_{\rm{layer}}}} \frac{||r||^2 dV}{V(\mathbbm{C}^{N_{\rm{layer}}})},
\end{aligned}
\end{equation}
where $r$ denotes the amplitude of the considered point,  $V(\mathbbm{C}^{N_{\rm{layer}}})$ denotes the volume of the whole space. It can be expected that the shape of a sphere could maximize the space utilization efficiency, resulting in the averaged power of
\begin{equation}
\label{power2}
\begin{aligned}
P_{\mathfrak{M}} = \frac{R^2}{N+1},
\end{aligned}
\end{equation}
where $R$ denotes the radius of the $2N$-sphere. With the normalized power constraint, we have $R=\sqrt{N+1}$. The upper bound of the minimum distance $d_{\rm{min}}$ can be approximated by
\begin{equation}
\label{d_min}
\begin{aligned}
a_{2N} d_{\rm{min}}^{2N} \leq \frac{a_{2N} R^{2N}}{M} =  \frac{a_{2N} (N+1)^{N}}{M},
\end{aligned}
\end{equation}
where $a_{2N} = \frac{\pi^N}{N!}$ denotes the coefficient for calculating the volume of $2N$-dimensional sphere. Finally, the minimum distance can be approximated by
\begin{equation}
\label{dis2}
\begin{aligned}
d_{\rm{min}} \leq \sqrt{N+1} M^{-\frac{1}{2N}}.
\end{aligned}
\end{equation}
It can be seen that both $\sqrt{N+1}$ and $M^{-\frac{1}{2N}}$ increase as the dimension $2N$ increases, proving the benefits for performing higher-dimensional modulation. As a comparison, 
if we consider a uniform constellation like QAM within the $2N$-dimensional cubic, the averaged power on each dimension can be written as
\begin{equation}
\label{power_qam}
\begin{aligned}
P_{\rm{QAM}} = 2\int_{0}^{R/2} \frac{r^2}{R} dr = \frac{R^2}{12}.
\end{aligned}
\end{equation}
With the normalized power as $1$ for every two dimensions, we have $R^2=6$, which is a constant invariant with $2N$. Since the volume of $2N$-dimensional cubic is $R^{2N}$, the corresponding minimum distance for QAM is written as
\begin{equation}
\label{dis_qam}
\begin{aligned}
a_{2N} d_{\rm{min,QAM}}^{2N} \leq \frac{R^{2N}}{M}.
\end{aligned}
\end{equation}
Recalling that the square of distance has impacted the result of error probability in (\ref{awgn_prob}), the ratio of minimum distances can be written as
\begin{equation}
\label{dis_qam_}
\begin{aligned}
\frac{d_{\rm{min}}^2}{d_{\rm{min,QAM}}^2} \approx \frac{N+1}{M^{\frac{1}{N}}} \cdot \frac{M^{\frac{1}{N}} a_{2N}^{\frac{1}{N}}}{6} = \frac{\pi (N+1)}{6 (N!)^{\frac{1}{N}}} \approx \frac{\pi e}{6},
\end{aligned}
\end{equation}
where the last approximation is obtained with $N! \simeq (N/e)^N$. These results are consistent with the ultimate shaping gain in [46]. To summarize, it can be seen that enlarging the modulation dimension can lead to an improved error probability and implicitly higher BICM capacity compared with legacy QAM constellation. 

Although it is attractive to deploy cross-layer modulation to further improve channel capacity, the design of high-dimensional constellation is extremely challenging, where AI/ML can be leveraged to break the impasse. Next we shall clarify the intrinsic relationship between BCE loss function and channel capacity.

Recalling the LLR $\hat{c}$ in (\ref{llr_ori}) calculated at receiver can be viewed as an estimation of $-{\rm{ln}} (l_1^{(i)}({\bf{y_s}}))$ in equation (\ref{llr}). Thus, the sigmoid of LLR $\hat{c}$ can be expressed as
\begin{equation}
\label{dis2}
\begin{aligned}
\tilde{c} &= \frac{1}{1+e^{-\hat{c}}} \approx \frac{1}{1+l_1^{(i)}({\bf{y_s}})}.
\end{aligned}
\end{equation}
Therefore, the BCE function can be further simplified as
\begin{equation}
\label{bce_simp}
\begin{aligned}
-\mathcal{L}_{\rm{bce}} & = c\log(\tilde{c}) + (1-c) \log(1-\tilde{c}) \\
& = -c \log (1+l_1^{(i)}({\bf{y_s}})) - (1-c) \log (1+l_0^{(i)}({\bf{y_s}})),
\end{aligned}
\end{equation}
which is exactly consistent with the second item of BICM capacity in (\ref{bicm_capacity2}). The theoretical analysis motivates us to use BCE as the loss function for the whole optimization problem, which coincides with the target of maximizing capacity from information theory perspective. Moreover, equation (\ref{awgn_prob}) indicates that the capacity is impacted by the channel condition. Therefore, the corresponding control mechanism is required to select the suitable modulation scheme. Moreover, the form in (\ref{awgn_prob}) is constrained by AWGN channel. For other channel models, the probability cannot be expressed in the close form, motivating us to explore AI-based control strategy to flexibly adapt to various channel conditions. 
}

\section{Simulation Results}\label{SectionIV}
In this section, numerical results of the proposed solutions are presented to quantify their performance gains against comparative schemes. The evaluation is conducted under 5G NR-compliant physical layer assumptions, adopting an integrated framework that accounts for both downlink and uplink transmissions, where data packets are transmitted over the downlink and CSI is conveyed via the uplink.

\begin{table}[t]
\centering
\caption{Basic Simulation Assumptions}
\label{tab1}
\setlength{\tabcolsep}{1.5mm}{
\begin{tabular}{|c|c|c|}
\hline
 & Parameters & Value   \\ \hline
\multirow{6}{*}{Uplink}         & Carrier frequency & 3.3GHz    \\ \cline{2-3}
 & Subcarrier spacing & 15kHz    \\ \cline{2-3}
 & Subcarrier number $N_{\mathrm{sc}}$ & 96    \\ \cline{2-3}
 & OFDM symbol number $N_{\mathrm{t}}$ & 1    \\ \cline{2-3}
 & Antenna number $N_{\mathrm{tx}} \times N_{\mathrm{rx}}$ & 1 $\times$ 32   \\ \cline{2-3}
 & Channel coding scheme & LDPC, Code rate=1/4,1/3,1/2   \\ \hline

\multirow{8}{*}{Downlink}         & Carrier frequency & 3.5GHz    \\ \cline{2-3}
 & Subcarrier spacing  & 15kHz    \\ \cline{2-3}
 & Subcarrier number $N_{\mathrm{sc}}$ & 144    \\ \cline{2-3}
 & Subband number $N_{\mathrm{sb}}$ & 3    \\ \cline{2-3}
 & OFDM symbol number $N_{\mathrm{t}}$ & 14    \\ \cline{2-3}
 & Antenna number $N_{\mathrm{tx}} \times N_{\mathrm{rx}}$ & 32 $\times$ 4 \\ \cline{2-3}
 & Payload size per RE  & 2,8,16,24,32 bits \\ \cline{2-3}
 & Channel coding scheme & LDPC, Code rate=1/2   \\ \hline

\multirow{3}{*}{Global}         & Channel model & CDL    \\ \cline{2-3}
 & UE speed & 3km/h    \\ \cline{2-3}
 & Channel estimation & Ideal    \\ \hline

\end{tabular}}
\end{table}


\begin{figure*}[tbp]
  \centering
  \small
  \subfloat[2 bits per RE]{
      \includegraphics[width=0.35\textwidth]{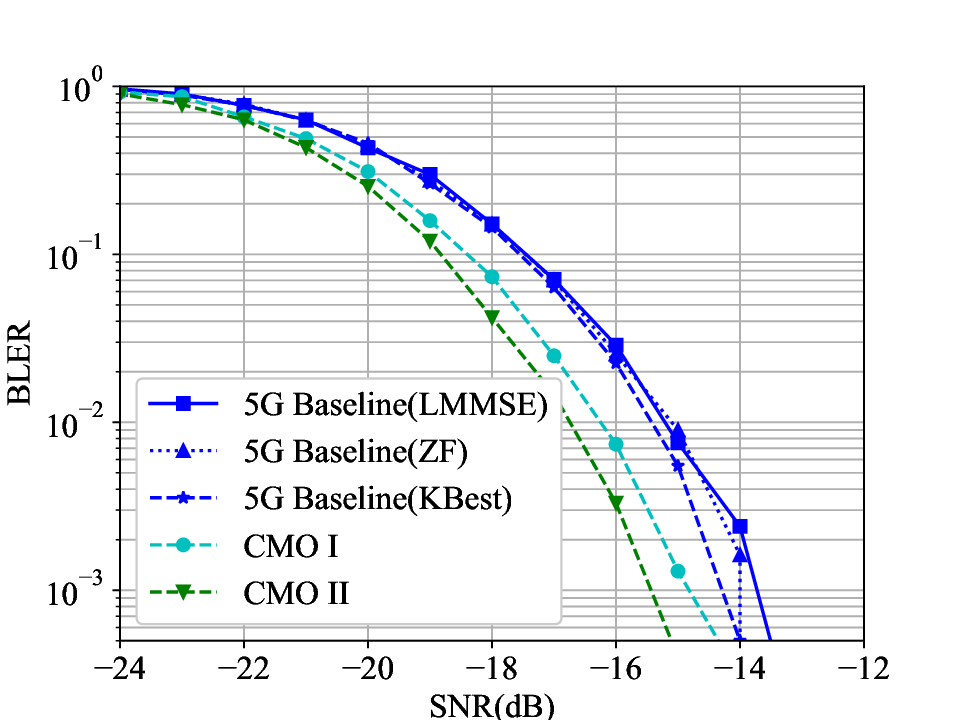}
\label{2bit1}
  }\hspace{-8mm}
  \subfloat[8 bits per RE]{
      \includegraphics[width=0.35\textwidth]{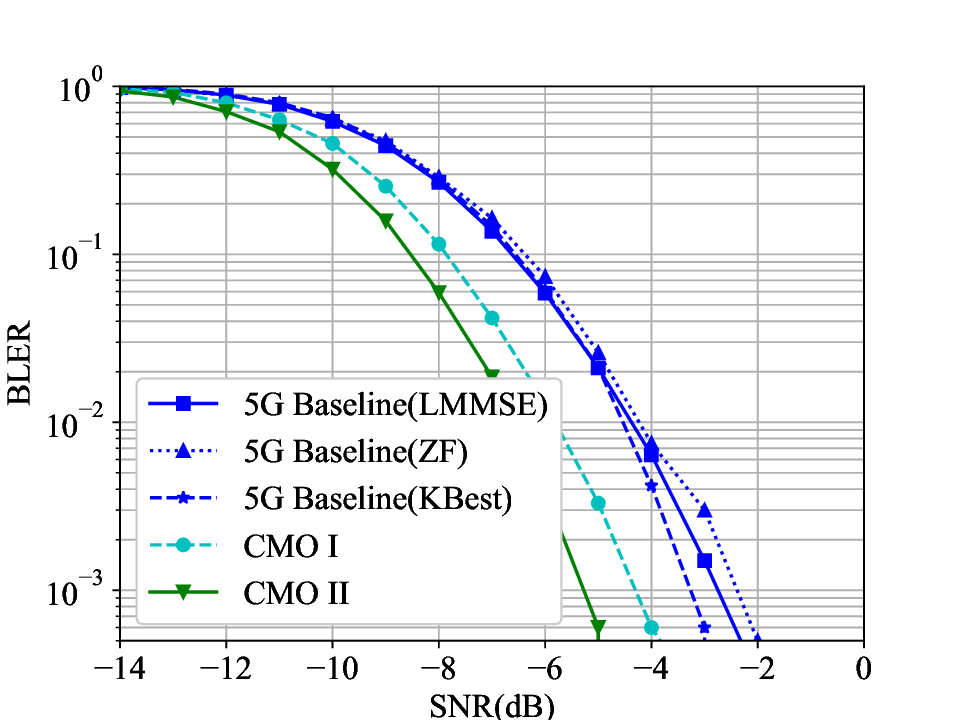}
  }\hspace{-8mm}
  \subfloat[16 bits per RE]{
      \includegraphics[width=0.35\textwidth]{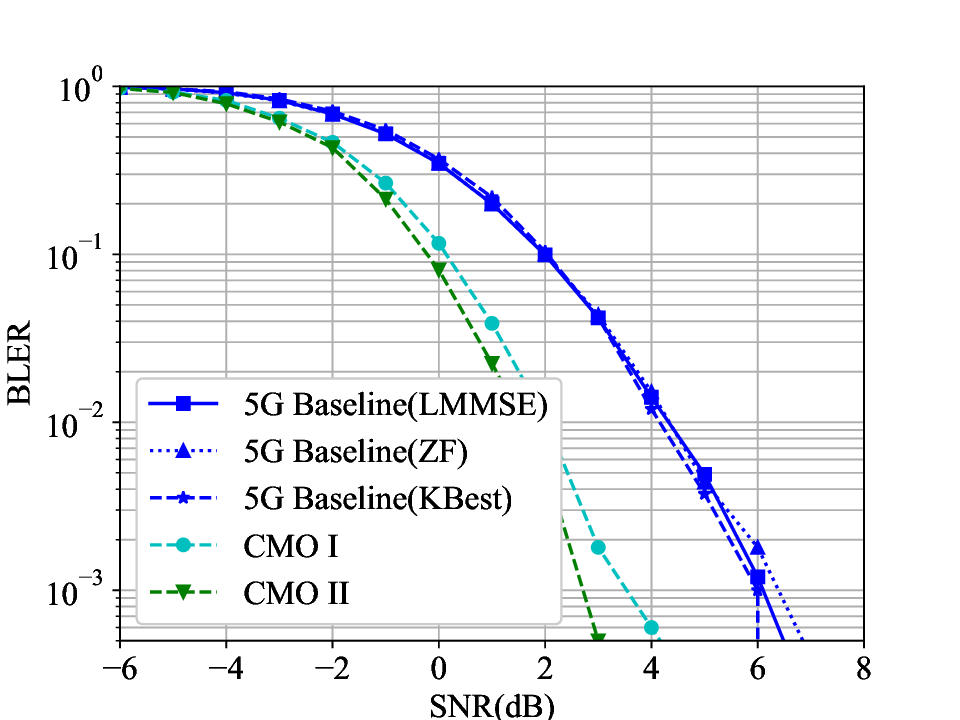}
\hspace{-3mm}
  }
  \\ \hspace{-8mm}
  \subfloat[24 bits per RE]{
      \includegraphics[width=0.35\textwidth]{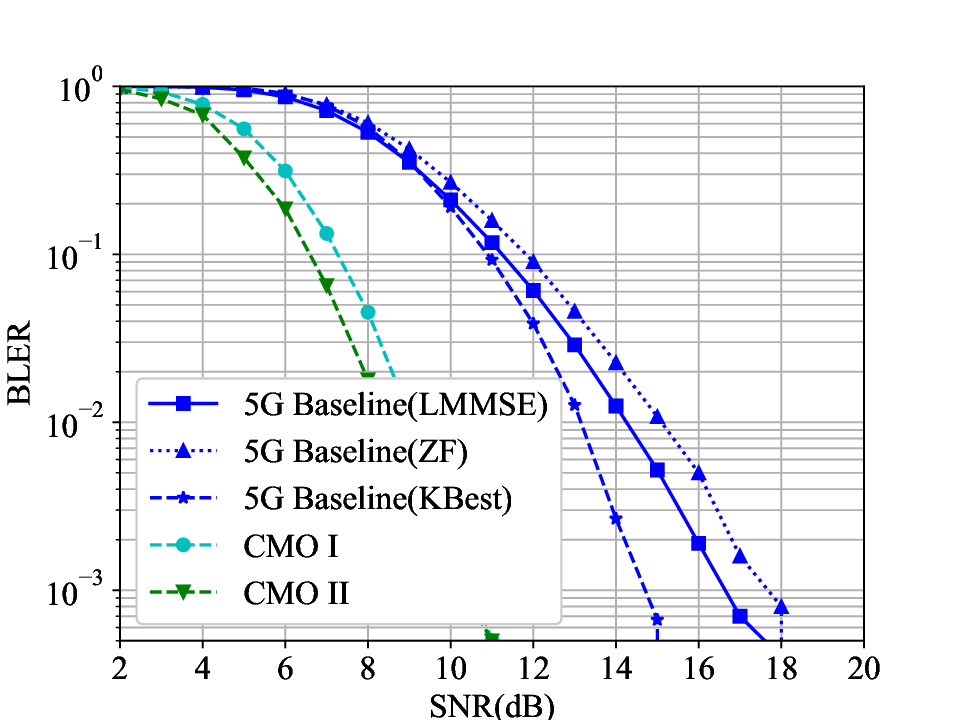}
  }\hspace{-8mm}
  \subfloat[32 bits per RE]{
      \includegraphics[width=0.35\textwidth]{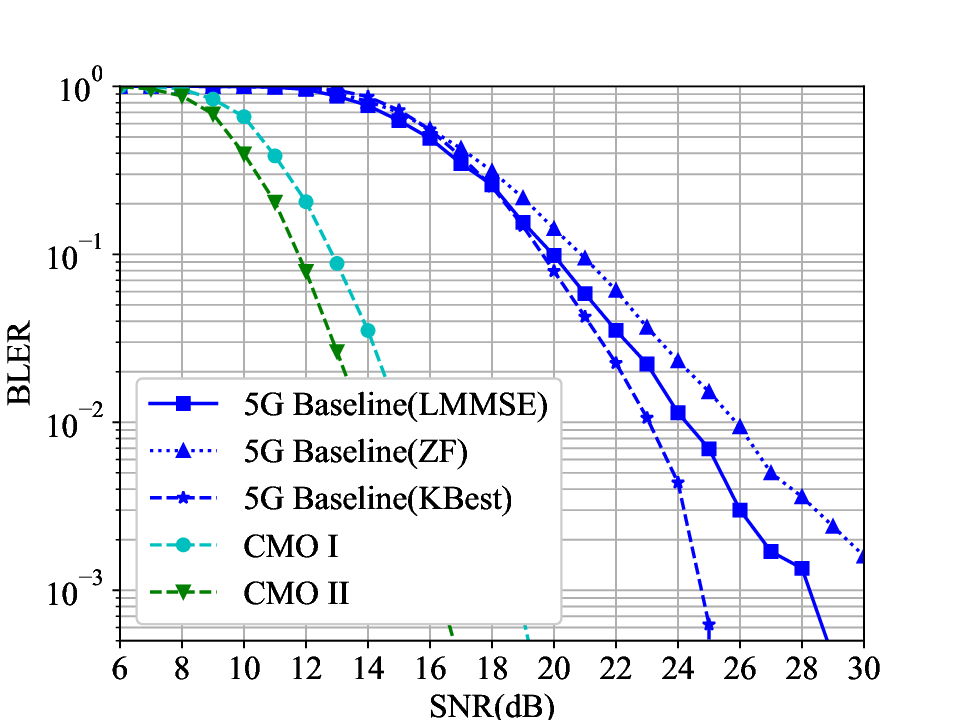}
\label{32bit1}
  }\hspace{-8mm}
  \subfloat[Link adaptation among all payload sizes]{
      \includegraphics[width=0.35\textwidth]{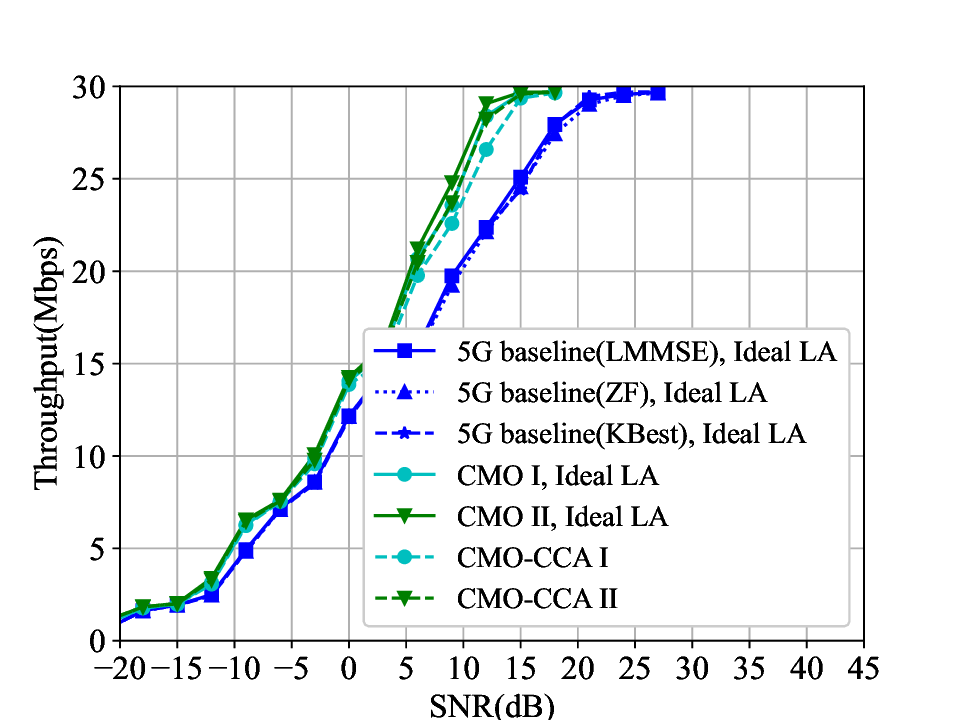}
\hspace{-3mm}
\label{TP1}
  }\hspace{-8mm}
\caption{Performance comparison under idealized uplink transmission}
\label{fig_sim1}
\end{figure*}

Some basic simulation parameters are summarized in Table \ref{tab1}. 
Specifically, the Clustered Delay Line (CDL)-C channel model, as defined by 3GPP and employed as a benchmark for link-level simulations, is adopted. To simplify the simulation, ideal channel estimation for both CSI acquisition and data reception is assumed in this section. 
Moreover, when idealized uplink transmission is assumed, all comparative schemes maintain a constant CSI feedback overhead of 192 bits, and the BS can obtain the bitstream without disturbance caused by air interface. By contrast, under practical uplink transmission assumptions, the CSI is transmitted across $N_{\mathrm{sc}}\times N_{\mathrm{t}}=$ 96$\times$1 $=$ 96 REs in the uplink resource grid, and the BS needs to process the signal that has passed through the uplink channel and construct the precoder accordingly.

To demonstrate the effectiveness of the proposed cross-module optimization (CMO) scheme, we evaluate two training variants, i.e., CMO I and CMO II, against several 5G-compliant baselines under identical system constraints. CMO I exclusively undergoes phase I training until convergence, while CMO II sequentially executes phase I followed by phase II training. Both CMO I and CMO II share identical model implementations and total training iterations but employ distinct training processes. The conventional comparative scheme, denoted as 5G baselines, utilize standardized constellations for modulation and demodulation, including quadrature phase shift keying (QPSK), 16 quadrature amplitude modulation (QAM), 64QAM, and 256QAM, and supports a configurable number of transmission layers ranging in $N_{\mathrm{layer}}\in\{1,2,3,4\}$. The eTypeII codebook-based CSI feedback is adopted where eigenvectors that are fed back at the BS serve as the precoding matrix. To comprehensively assess baselines performance, three classical detection algorithms are employed. The LMMSE receiver aims to minimize the mean square error, thereby balancing interference suppression with noise enhancement. In contrast, the Zero-Forcing (ZF) receiver focuses solely on complete interference cancellation, which often leads to significant noise amplification. As a non-linear alternative, the K-Best receiver, a sub-optimal sphere decoding algorithm, provides a favorable performance-complexity trade-off by approximating the Maximum Likelihood (ML) performance. To focus solely on the downlink data transmission performance gains of the proposed CMO schemes, idealized uplink transmission conditions are assumed first. The CSI encoder and decoder components of both CMO I and CMO II retain the original EVCsiNet-T \cite{han2024ai} architecture to ensure compatibility with CSI feedback overhead in bitstream form.

As shown in Fig. \ref{2bit1} to Fig. \ref{32bit1}, the BLER performance of CMO I, CMO II and 5G baselines is evaluated across different payload sizes. 
For the 5G baselines, optimal link adaptation is achieved by exhaustively evaluating all feasible modulation and layer combinations for each transmission and selecting the one minimizing BLER where the combinations adapt the corresponding payload sizes. In contrast, CMO I and CMO II assume a fixed maximum layer count of $N_{\mathrm{layer}}=4$ for all transmissions, since the number of transmission layers can be indirectly adjusted by cross-layer modulation or precoder. 
First, Fig. \ref{2bit1} to Fig. \ref{32bit1} reveal that the CMO I consistently outperforms the 5G baselines across all considered payload sizes. This implies the performance gain achieved by the proposed AI/ML cross-layer modulation. 
Moreover, under identical model implementations, CMO II consistently outperforms CMO I across all payload sizes, with performance gain attributed to utility-oriented precoder construction. 
Furthermore, it can also be noticed that the performance gaps between proposed CMO II and 5G baselines widen as the payload increases. For instance, at a BLER of 0.1, the gain evolves from around 1 to 2 dB in the low-bit region to surpassing 5 dB in the high-bit region. This trend highlights that as payload sizes increase and signal constellations grow more intricate, the advantages of the proposed AI/ML-based cross-module optimization become increasingly evident. 
The throughput comparison under various SNRs is depicted in Fig. \ref{TP1}.  Ideal link adaptation (Ideal LA), serving as a theoretical benchmark, is applied to both baselines and CMO schemes under comparison. 
For AI/ML-based schemes of CMO I and CMO II, performance under the proposed cooperative control agents (CCA) is further assessed. The results demonstrate that under the proposed CCA, the throughput performance of both CMO I and CMO II closely approximates their ideal link adaptation counterparts, where the superiority of the proposed solution is demonstrated and established in practical SNR conditions. 
When compared to the 5G baselines with ideal link adaptation, the CMO-CCA I scheme demonstrates a throughput gain of approximately 5\% at SNR = 0dB and 10\% at SNR = 10dB. Similarly, the CMO-CCA II outperforms the 5G baselines by about 7\% at SNR = 0dB and 16\% at SNR = 10dB, where the advantages of the proposed solutions are more comprehensively explained.

\begin{figure*}[tbp]
  \centering
  \small
  \subfloat[2 bits per RE]{
      \includegraphics[width=0.35\textwidth]{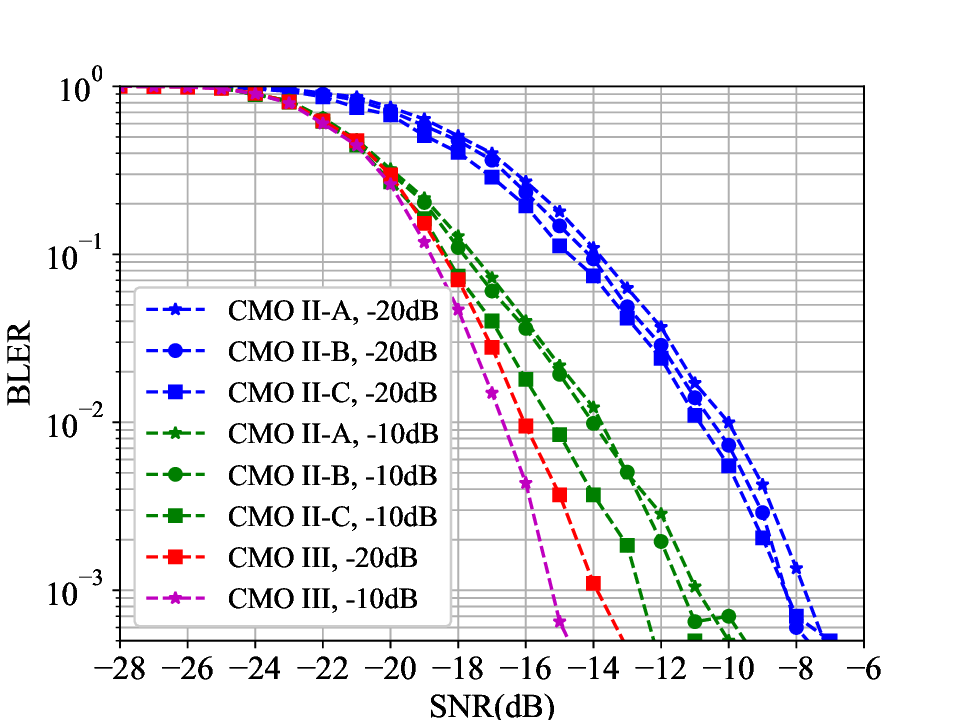}
\label{2bit2}
  }\hspace{-8mm}
  \subfloat[8 bits per RE]{
      \includegraphics[width=0.35\textwidth]{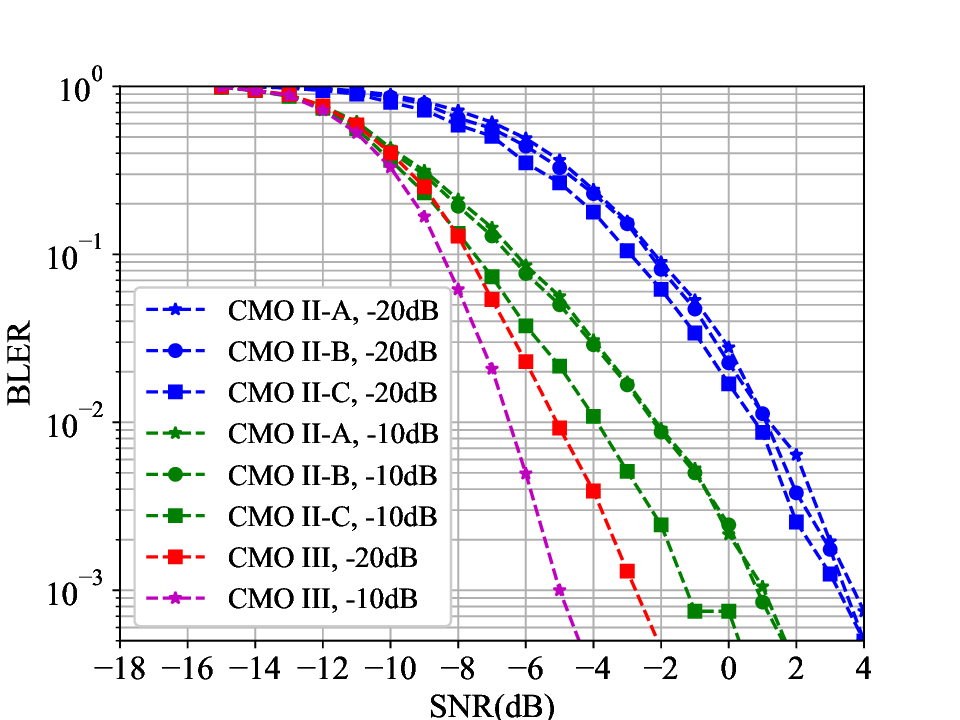}
  }\hspace{-8mm}
  \subfloat[16 bits per RE]{
      \includegraphics[width=0.35\textwidth]{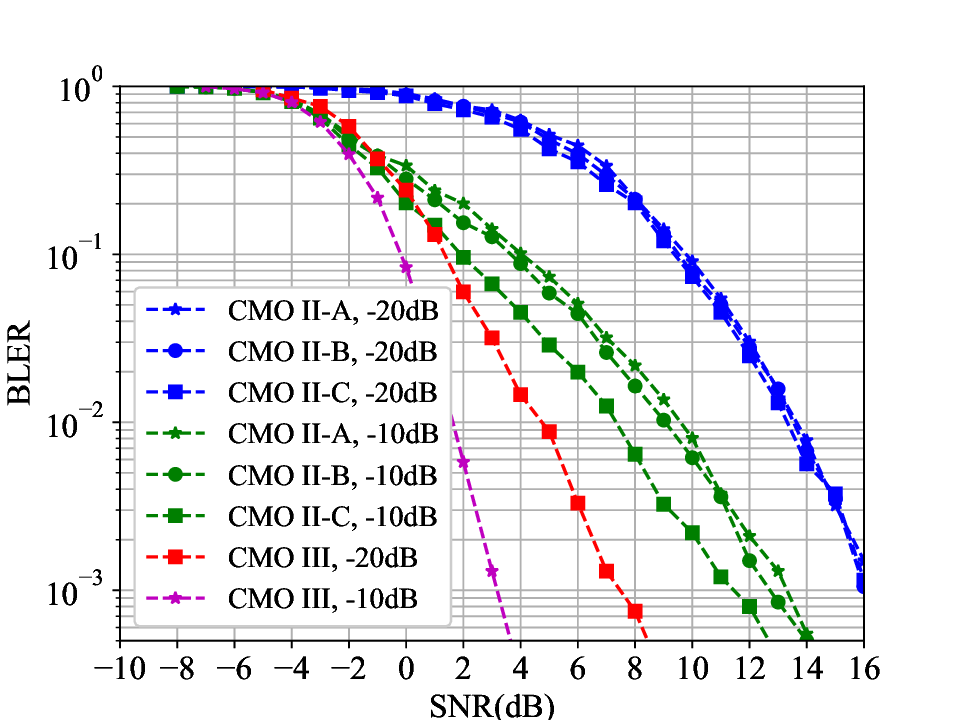}
\hspace{-3mm}
  }
  \\ \hspace{-8mm}
  \subfloat[24 bits per RE]{
      \includegraphics[width=0.35\textwidth]{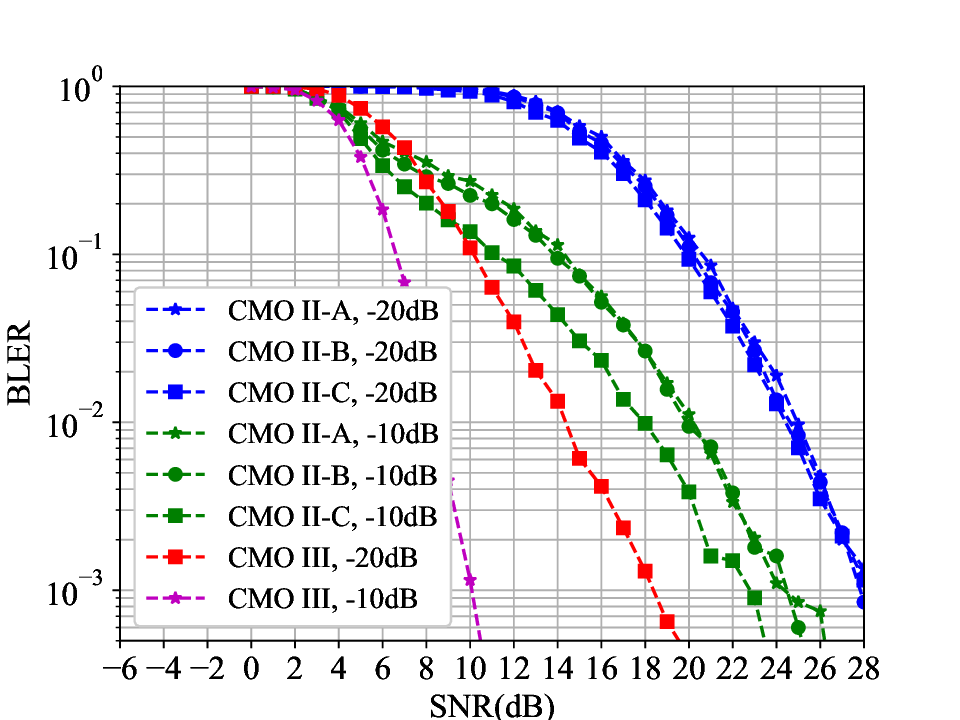}
  }\hspace{-8mm}
  \subfloat[32 bits per RE]{
      \includegraphics[width=0.35\textwidth]{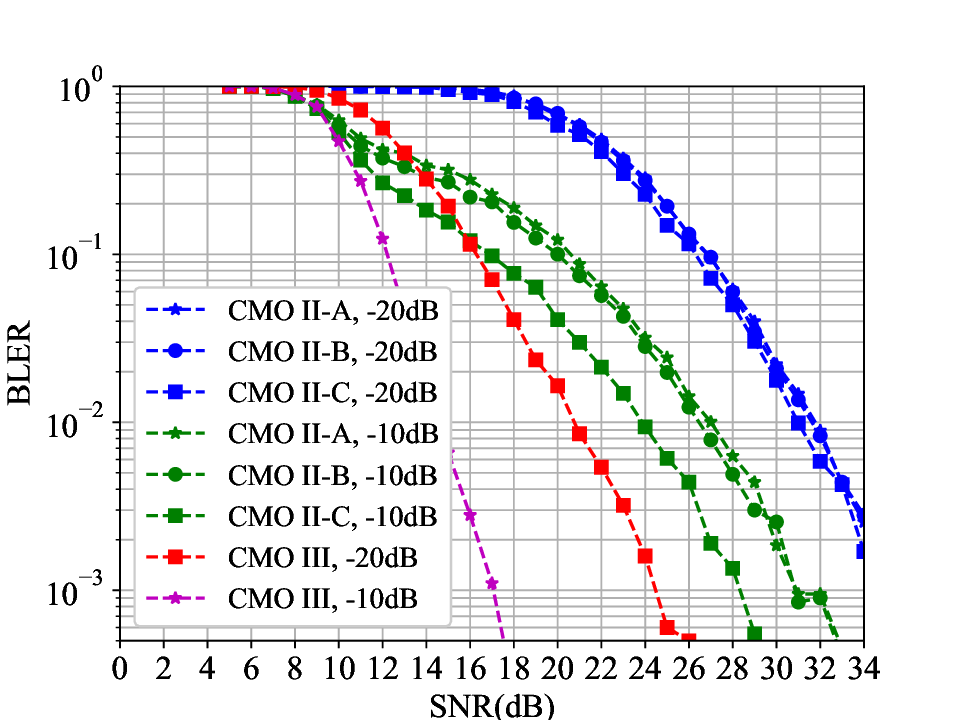}
\label{32bit2}
  }\hspace{-8mm}
  \subfloat[Link adaptation among all payload sizes]{
      \includegraphics[width=0.35\textwidth]{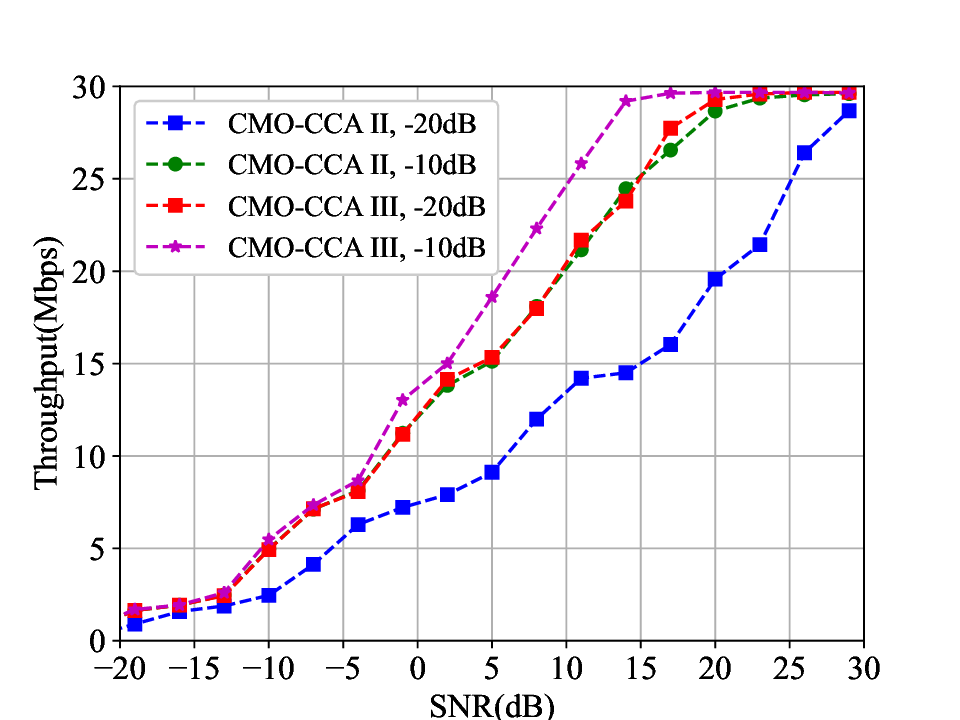}
\hspace{-3mm}
\label{TP2}
  }\hspace{-8mm}
\caption{Performance comparison under practical uplink transmission}
\label{fig_sim2}
\end{figure*}

\begin{figure}[tbp]
\setlength{\abovecaptionskip}{0.cm}
\setlength{\belowcaptionskip}{-0.cm}
\centering
\includegraphics[width=0.95\linewidth]{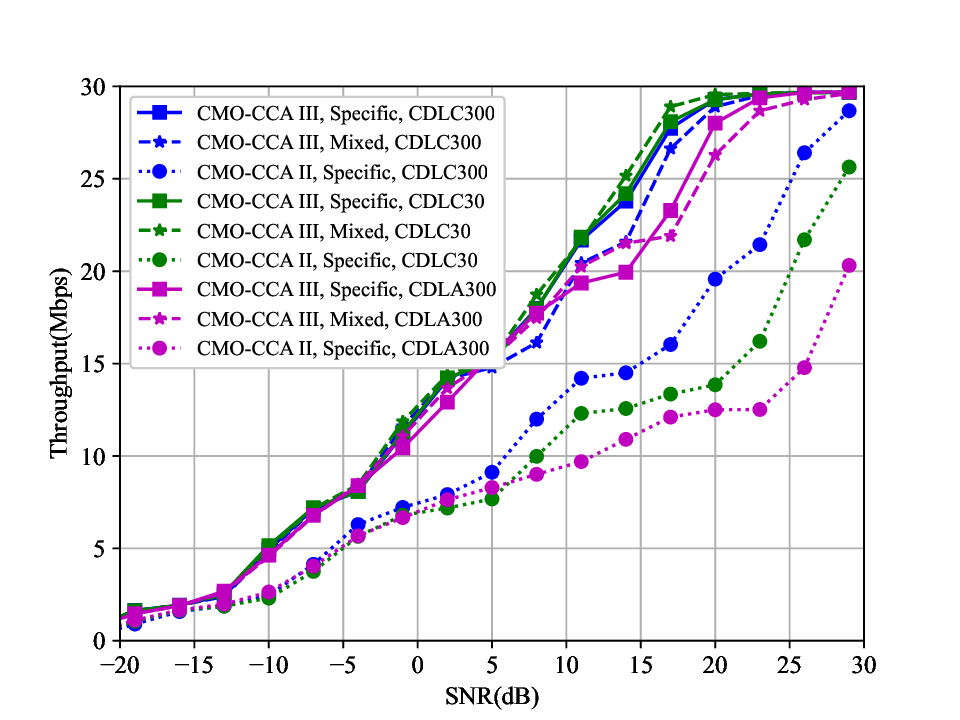}
\caption{Generalization study of proposed scheme for different channel models}
\label{fig_sim3}
\end{figure}

To demonstrate the effectiveness of the proposed CSI feedback enhancement to adapt to signal distortions introduced by practical uplink transmission, we further compare the CMO II with CMO III. Beyond CMO II, the CMO III further upgrades the bit-level bottlenecks bypassing for CSI feedback, as described in subsection \ref{bitbypass}. 
In practical communication scenarios, uplink transmission often suffers from poorer SNR than downlink transmission due to UE power constraints. To comprehensively evaluate the proposed scheme under such limited uplink conditions, this part considers challenging uplink SNRs of -20 dB and -10 dB, mimicking realistic power-constrained environments. 
For comparison, the CMO II is directly applied to practical uplink transmission process. Specifically, the bitstream output by its CSI encoder undergoes 5G-compliant channel coding, modulation, and mapping to $N_{\mathrm{sc}}\times N_{\mathrm{t}}=$ 96$\times$1 $=$ 96 REs. After experiencing actual channel transmission, the CSI-carrying signals go through channel equalization, demodulation, and channel decoding. The recovered bitstream is then fed into the CSI decoder to construct the precoder. 
Similar to downlink scenarios, the uplink transmission of CMO II can consider different combinations of channel coding rates and modulation orders to map the 192-bit output of the CSI encoder onto 96 REs. Three potential configurations to balance spectral efficiency and error resilience are evaluated, i.e., coding rates and modulation orders of 1/4 with 256QAM, 1/3 with 64QAM, and 1/2 with 16QAM, denoted as CMO II-A, CMO II-B, and CMO II-C, respectively.

As depicted in Fig. \ref{2bit2} to Fig. \ref{32bit2}, the BLER performance across varying downlink payload sizes is compared between the CMO II and CMO III approaches. The CMO III demonstrates consistent gains across all payload configurations, where the gain increases as the payload size grows. 
Notably, the CMO III scheme achieves more pronounced advantages under poorer uplink signal conditions. At an uplink SNR of -10 dB, taking BLER = 0.1 as a reference, the performance gain over the CMO II grows from approximately 1 dB in low-bit regions to nearly 5 dB in high-bit regions. In more severe scenarios with uplink SNR of -20 dB, the disparity widens significantly. That is, the gain escalates from around 4 dB at 2 bits per RE to up to 10 dB at 32 bits per RE under the same BLER threshold of 0.1. These results clearly demonstrate that the design enhancement for CSI feedback in CMO III effectively boosts communication performance in scenarios with limited uplink link quality. 
As illustrated in Fig. \ref{TP2}, the throughput performance of both CMO II and CMO III under CCA is compared across varying signal conditions. Consistent with the trends observed in the BLER results, the CMO III exhibits more substantial throughput gains at lower uplink SNR, highlighting its robustness in UE power-constrained situation. At an uplink SNR of -10 dB, the CMO III achieves a throughput gain of approximately 22\% over the CMO II when the downlink SNR is 10 dB. In contrast, under a more challenging uplink SNR of -20 dB, the gain increases to around 52\% at the same downlink SNR level. These further explain the advantages of the proposed solutions.

To further assess the generalization capabilities of the proposed schemes, Fig. \ref{fig_sim3} presents performance comparison under diverse wireless environments. The CMO-CCA III is trained using mixed datasets containing both CDL-A and CDL-C channel with distinct delay spreads of 30 ns and 300 ns (denoted as Mixed), and then tested on corresponding target channels. For comparative analysis, specialized implementations of both CMO-CCA III and CMO-CCA II (denoted as Specific) are exclusively trained and tested on individual target channel datasets. Experimental results show that the mixed-trained CMO-CCA III achieves throughput performance comparable to its specifically trained counterpart, demonstrating marginal performance variation despite broader training scope. Notably, the mixed-trained CMO-CCA III outperforms the specifically trained CMO-CCA II in throughput metrics. These observations collectively validate the excellent generalization capacity of the proposed scheme when dealing with different channel conditions.

\begin{figure}[tbp]
\setlength{\abovecaptionskip}{0.cm}
\setlength{\belowcaptionskip}{-0.cm}
\centering
\includegraphics[width=0.95\linewidth]{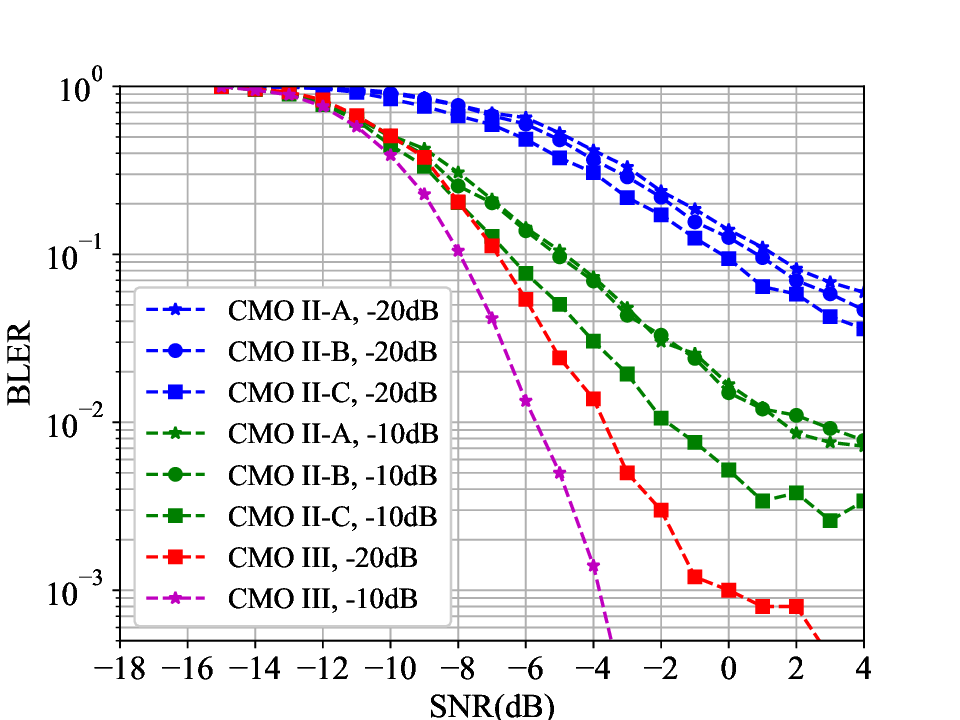}
\caption{Performance comparison under multiple BSs and multiple UEs scenario}
\label{fig_sim4}
\end{figure}

To rigorously validate the robustness of the proposed CMO III under more realistic and challenging conditions, we extend the evaluation to a scenario involving two BSs and two UEs coexisting. For downlink data transmission, the two BSs transmit signals over the same time-frequency resources, leading to mutual inter-BS interference. For uplink feedback, the two UEs also share identical resources for signal transmission, resulting in mutual inter-UE interference. In this experimental setup, the target-to-interference power ratio is set to 1/0.3 to mimic practical interference conditions. As illustrated in Fig. \ref{fig_sim4}, under a payload size of 8 bits per RE, CMO III still maintains a significant performance gain over CMO II. Notably, the gain of CMO III over CMO II is more pronounced at a UL SNR of -20 dB than at -10 dB, aligning with the general trend observed in the interference-free scenario. Comparing Fig. \ref{fig_sim4} with the interference-free scenario in Fig. \ref{fig_sim2} reveals that CMO III exhibits minimal performance degradation, significantly smaller than that of CMO II. This result confirms that the end-to-end training under interference conditions enhances the overall interference tolerance at both the transmitter and receiver ends. More specifically, the enhanced CSI feedback mechanism in CMO III provides superior robustness against uplink interference, making it less affected by such interference than CMO II.

The computational and storage complexity are analyzed through two key metrics: floating point operations (FLOPs) and the count of trainable parameters. Regarding uplink transmission components, the CSI encoder and decoder model demonstrate architecturally symmetric designs, consequently exhibiting identical computational requirements with 10.598 million FLOPs and 1.863 million parameters each. For downlink transmission components, the computational demands of the cross-layer modulation and demodulation model per RE are detailed in Table \ref{tab2}. While both FLOPs and parameter counts show dependence on transmission payload sizes, this variation remains relatively negligible when compared to the intrinsic complexity of the model architecture itself. Notably, especially for the low-payload transmission, there exists significant potential to substantially reduce the model complexity of cross-layer modulation and demodulation without compromising performance. Finally, the complexity introduced by the CCA model for enabling dynamic switching between transmission schemes is proven to be minimal, requiring only 0.037 million FLOPs and 0.018 million parameters.

\begin{table}[tbp]
\centering
\caption{Evaluation of FLOPs and number of trainable parameters of cross-layer modulation and demodulation model}
\label{tab2}
\renewcommand\arraystretch{1} 
\setlength\tabcolsep{4pt} 
\begin{tabular}{|c|c|c|c|c|}
\hline
\multirow{2}{*}{Bits per RE}    &\multicolumn{2}{c|}{FLOPs ($\times 10^6$)}  &\multicolumn{2}{c|}{Parameters ($\times 10^6$)}  \\  \cline{2-5}
& Modulator & Demodulator & Modulator & Demodulator  \\ \hline
2   &   0.398   &   2.140   &   0.200   &   0.533   \\ \cline{1-5}
8   &   0.401   &   2.146   &   0.202   &   0.535   \\ \cline{1-5}
16   &   0.406   &   2.154   &   0.204   &   0.537   \\ \cline{1-5}
24   &   0.410   &   2.163   &   0.206   &   0.539   \\ \cline{1-5}
32   &   0.414   &   2.171   &   0.208   &   0.541   \\ \cline{1-5}
\end{tabular}
\end{table}

\section{Standardization Discussion}\label{sectionV_sub}
6G is characterized by a renewed opportunity to architect a systemic transformation of communication systems. In this work, leveraging the transformative potential of AI, we focus on cross-layer and cross-module joint design solutions, and propose the introduction of dedicated control agents at both the UE and network sides to comprehensively manage various wireless AI solutions. Considering the process from scheme design to actual implementation, the analysis of standardization impacts is further provided.
\begin{itemize}
\item 
The proposed cross-module optimization schemes may introduce new requirements for layer mapping procedure, MCS and CQI feedback design. For instance, the input code block should be jointly modulated to multiple layers, instead of isolated modulation and layer mapping in 5G NR. Moreover, when considering high-dimensional cross-layer modulation, further research and standardization efforts are needed in areas such as the joint feedback for constellation diagram and rank, as well as potential modulation enhancements like probabilistic shaping.
\item 
The introduced control agent represents an AI-based LCM solution. Since 3GPP Release 18, LCM for wireless AI solutions has been a key focus in standardization. However, during the 5G phase, LCM discussions primarily centered on non-AI solutions, which aimed to manage AI solutions through standardized model identification, monitoring, and switching. Relying solely on these 5G-era approaches for model management and two-sided alignment would inevitably result in significant signaling overhead and impose substantial standardization burdens. The introduction of control agents facilitates more intelligent LCM management. However, it should be noted that the issue of inter-vendor collaboration between control agents also requires further consideration in future with standardization efforts.
\end{itemize}

\section{Conclusion}\label{sectionV}
This article proposes a framework of AI-native cross-module optimized physical layer with cooperative control agents, encompassing holistic optimization across physical layer AI/ML modules integrated with multiple enhancement mechanisms and control strategies. Specifically, simultaneous optimization across global modules breaks down traditional inter-module information silos to facilitate end-to-end training toward global objectives. Moreover, AI/ML-based cross-layer modulation breaks the isolation of inter-layer features, thus expanding the constellation solution space. The theoretical analysis that reveals the rationality of cross-layer modulation is also provided. The utility-oriented precoder construction directly generates the precoder optimized for end-to-end performance. By integrating modulation into AI/ML-enabled CSI feedback, the bit-level bottlenecks are effectively bypassed. Additionally, AI/ML-based control agents facilitate global throughput optimization with integrated control mechanisms. Finally, simulation results using practical 3GPP assumptions demonstrate the superiority of the proposed solutions in terms of block error rate and throughput. Above contents of this article also hopefully provide valuable insights for 3GPP discussions in the future. In the future work, to maintain scalability, several approaches expanded from our proposed framework can be considered, such as subband-level model processing partial bandwidth segments sequentially and pretraining a base model offline followed by efficient online fine-tuning with limited samples, etc. We also recognize the importance of comprehensive channel estimation robustness in practical deployments, and are actively investigating AI-native solutions for more practical scenarios including ICI and ISI mitigation as a direction for further research.

\section*{Acknowledgment}
The authors would like to thank Dr. Zidong Wu for his outstanding support and valuable contributions to this work, particularly his profound insights and rigorous efforts in the theoretical derivation and analysis.

\bibliographystyle{IEEEtran}
\bibliography{IEEEabrv,ref}

\vfill

\end{document}